\pdfoutput=1
\documentclass{sig-alternate}

\usepackage{times,amsmath,amssymb}
\usepackage{framed}
\usepackage{stmaryrd}
\usepackage{listings}
\usepackage{cite}
\usepackage{bussproofs}
\usepackage[framed,amsmath,thmmarks]{ntheorem}
\usepackage{color}
\usepackage{listings}
\usepackage{graphicx}
\usepackage{url}
\usepackage[pdftex]{hyperref}

\newenvironment{myitemize}{%
\begin{list}{$\bullet$}{\topsep3pt\parskip0pt\partopsep0pt\itemsep-1pt\labelwidth1pt\labelsep3pt\leftmargin8pt}}%
{\end{list}}
 {\endMakeFramed}

\newtheorem{lemma}{Lemma}[section]

\newtheorem{theorem}{Theorem}[section]

\theorembodyfont{\upshape}
\newtheorem{algorithm}{Algorithm}[section]
\newtheorem{definition}{Definition}[section]

\theorembodyfont{\upshape}
 \newtheorem{example}{Example}[section]

\newcommand{\DOM}[1]{\ensuremath{\mathcal{D}\textit{om}(#1)}}

\newcommand{\LANG}[1]{\ensuremath{\mathcal{L}(#1)}}

\newcommand{\defaultAuto}[1][]{\ensuremath(\Sigma,Q#1,\mathcal{T}#1,\mathcal{B}#1,\mathcal{S}#1,\delta#1)}
\newcommand{\Run}[2]{\ensuremath R^{#2}_{#1}}

\newcommand{\reach}[3]{\ensuremath #2 \rightarrow_{#1} #3 }
\newcommand{\reachstar}[3]{\ensuremath #2 \rightarrow_{#1}^{*} #3 }

\iftrue

\fi

\title{
XPath Whole Query Optimization}
\numberofauthors{2}
\author{
\alignauthor
Sebastian Maneth\\
       \affaddr{NICTA and UNSW}\\
       \affaddr{Sydney, Australia}\\
       \email{sebastian.maneth@nicta.com.au}
\alignauthor
Kim Nguyen\\
       \affaddr{NICTA}\\
       \affaddr{Sydney, Australia}\\
       \email{kim.nguyen@nicta.com.au}
}
\begin{document}
\maketitle
\begin{abstract}
Previous work reports about SXSI, a fast XPath engine 
which executes tree automata over compressed XML indexes.
Here, reasons are investigated why SXSI is so fast.
It is shown that tree automata can be used as a general framework
for fine grained XML query optimization. 
We define the ``relevant nodes'' of a query
as those nodes that a minimal automaton must touch in order to answer
the query. 
This notion allows to skip many subtrees during execution,
and, with the help of particular tree indexes, even allows to
skip internal nodes of the tree. 
We efficiently approximate runs over relevant nodes
by means of on-the-fly removal of alternation and non-determinism
of (alternating) tree automata.
We also introduce many implementation techniques which allows us to efficiently
evaluate tree automata, even in the absence of special indexes.
Through extensive experiments, we demonstrate the impact of the different
optimization techniques.
\end{abstract}

\section{Introduction}
\label{sec:introduction}
The XPath query language plays a central role in XML processing:
it is deeply uprooted in almost every XML technology, starting
from query languages such as XQuery and XSLT, to
access control languages such as XACML, to JavaScript engine of popular 
web browsers.
Thus, efficient XPath evaluation is essential 
for any time-critical XML processing.
In this paper we show how tree automata can be
used as framework for fine-grained and novel types of 
XPath query optimizations.
The experiments with our prototype show that,
together with appropriate indexes for the XML document tree,
these optimizations give rise to unprecedented execution
speed for XPath queries, outperforming the
fastest existing XPath engines.\looseness-1

The first breakthrough in efficient XPath execution was
Koch \emph{et al.}'s seminal paper~\cite{DBLP:conf/vldb/GottlobKP02} (see
also~\cite{DBLP:journals/jacm/GottlobKPS05}) 
where it is shown
that Core XPath can be evaluated in time $O(|D|\cdot|Q|)$ where $|D|$ is the size of the document and
$|Q|$ is the size of the query.
Core XPath refers to the tree navigational fragment of XPath.
Considering the 
time bound of Koch's algorithm, there
are two obvious ways of reducing this complexity in practice:
\begin{myitemize}
\item[\textbf{(1)}] reduce the number of query steps (``$|Q|$-optimization'') and
\item[\textbf{(2)}] reduce the number of nodes to consider (``$|D|$-optimization'').
\end{myitemize}

{\bf Extreme $|Q|$-Optimization:} 
A top-down deterministic tree automaton (TDTA) processes an input tree
starting in its initial state, at the root node. 
It then applies a unique rule which says, 
for a given state and label of a node, how
to process the children of that node.
A node is selected as a result, if the unique state reached by the
automaton on that node and the label of that node are elements of a special
``set of selection pairs''.
After compiling a (restricted) XPath query into such an automaton
(which takes $O(|Q|)$ time), the run function 
only requires a single look-up at each node of the input tree
(plus possibly an insertion of the current node into 
the result list. Since the function visits the nodes in document order
and only once, this insertion can be performed in constant time, keeps
the result sorted and duplicate-free). Thus, the
evaluation runs in $O(|D|)$ time,
giving the extreme case of $|Q|$-optimization to $|Q|=1$.
Similar automata for XML processing have been 
considered~\cite{DBLP:journals/tcs/NevenS02,
  DBLP:conf/fsttcs/NeumannS98,DBLP:conf/dbpl/NiehrenPTT05}.  
However, implementations of such automata cannot compete with 
state-of-the-art XPath engines.
The reasons for this deficiency are that (1) performance depends on
the speed of firstChild and
nextSibling operations in the XML tree data structure, (2) the automaton needs to visit \emph{every} node of $D$
and (3) the compilation into TDTA only works for a very restricted
  subset of Core XPath. 

To address (1), many implementations use in-memory pointer structures.
However, this blows up the memory requirement by a factor of 5-10 over
the size of the original XML document. Hence, such implementations
can only work over small documents.
We solve this problem by using state-of-the-art
succinct trees~\cite{DBLP:journals/corr/abs-0905-0768}, a recent
development in data structures.

Solutions to problems (2) and (3) are the main subject of this paper.
We study ways to restrict the nodes of the document which must be
visited by the run function of the automaton. 
This gives rise to the notion
of \emph{relevant nodes}, one of our key contributions.
To address (3), we work with non-deterministic alternating tree automata
and carefully develop on-the-fly determinization and alternation elimination
algorithms. This allows to retain most 
the benefits of deterministic automata
while increasing the expressive power to full Core XPath.
Altogether, our implementation of these solutions to (1) -- (3) provides 
XPath execution speed competitive with the best
known engines~\cite{DBLP:journals/corr/abs-0907-2089}.
While we restrict ourselves for didactic reasons to a fragment of Core
XPath, our prototype ``SXSI'' implements Core XPath plus
text predicates~\cite{DBLP:journals/corr/abs-0907-2089}; we
are currently adding other XPath 1.0 features such 
as number functions and aggregates.

{\bf $|D|$-Optimization using Relevant Nodes}
Consider the query $Q_0=\text{//{\tt a}//{\tt b}}$ 
which selects all {\tt b}-descendants of {\tt a}-labeled nodes.
A TDTA for this query starts at the root in a state $q_0$.
When it encounters an {\tt a}-node it changes to a state $q_1$.
Any {\tt b}-node encountered in $q_1$ is selected as result.
For such an automaton we say that a node is \emph{relevant},
whenever the automaton changes state, or selects a node.
Thus, all top-most {\tt a}-nodes  and all their 
{\tt b}-labeled descendants are relevant. Note that for this query, 
one could use the staircase join~\cite{DBLP:conf/vldb/GrustKT03} to
restrict the set of all {\tt a}-nodes 
to the top-most ones, and only then select {\tt b}-descendants; in this way
only the relevant {\tt b}-nodes are touched (but some non-relevant
{\tt a}-nodes might be touched in the first step).
Here, we first give an algorithm that executes an arbitrary
TDTA so that \emph{only relevant nodes are visited}.
This is achieved by executing the automaton over an index
that allows at any node
to ``jump''  to the next $\sigma$-labeled descendant
(for any label $\sigma$) or to the next $\sigma$-labeled following node
(according to XPath), for any $\sigma$. 
For bottom-up deterministic tree automata (BDTA), we can define
relevant nodes in a similar way. We sketch an algorithm for BDTAs
that \emph{only touches relevant nodes}, given an index that 
allows access to all bottom-most nodes with a given label 
and allows to jump to labeled ancestors (due to space constraint and
the fact that the bottom-up algorithm has to handle more cases than
the top-down one to ensure that nodes are only visited once, we do not
give it fully in this paper). 

Given a query, it is not always possible to determine which one of the
bottom-up or top-down evaluation is the most efficient (\emph{i.e.} visits fewer
nodes). For instance, for query $Q_0$, if the input document has
less {\tt b} nodes than {\tt a} nodes, a bottom-up traversal seems
more efficient. Following this idea, we extend our evaluation
algorithm to support \emph{Start Anywhere Runs}: for a query such as
//{\tt a}//{\tt b}//{\tt c}, if the global count of {\tt  b}-nodes is low, we can jump
to these {\tt b}-nodes, and from there execute simultaneously a
bottom-up run which checks for {\tt a}-nodes and a top-down run which selects {\tt
  c}-nodes.

\textbf{Non-Deterministic Automata}
To determine the relevant nodes for a TDTA or BDTA, we actually
first have to \emph{minimize} the automaton.
Intuitively, a non-minimal automaton can do many useless state-changes.
While minimization can efficiently be done for deterministic automata,
it poses a big problem for non-deterministic automata. Here,
minimization is EXPTIME-complete, and, there need not even exist
a unique minimal automaton. 
Unfortunately, for XPath 
we \emph{must} deal with non-deterministic automata:
consider $Q_1=\text{//{\tt a}[.//{\tt b}]//{\tt c}}$.
If we execute it top-down and are below an {\tt a}-node, then 
for a {\tt c}-node we cannot know whether to select it (this
depends on the presence of {\tt b}-nodes which might be below).
Similarly, the query //{\tt a}//{\tt c} cannot be done in a
deterministic bottom-up way. 
There is an elegant way to characterize relevant nodes for
non-deterministic automata, using equivalence between sub-automata.

This notion proves too complex to implement in practice (equivalence
is EXPTIME-complete), but we give an \emph{on-the-fly} algorithm which
soundly
\emph{approximates} the relevant nodes of a nondetermi\-nistic 
tree automaton, while evaluating the automaton on an input tree.
Our experiments show that for typical XPath queries our
on-the-fly algorithms perform well: the approximation of the set of
relevant nodes that we compute is close to the real set allowing us to
only visit a small fraction of the complete document.

\textbf{Plan} Section~\ref{sec:auto} gives the definitions and
introduces our model of selecting tree
automata. Section~\ref{sec:relnodes} formally defines the concept of
relevant nodes and studies two optimal algorithms for minimal top-down
and bottom-up selecting tree automata. Section~\ref{sec:implem}
introduces our variant of
alternating tree automata, their encoding of XPath
queries, and presents the approximating algorithm as well as a collection of
implementation techniques. The impact
of these  techniques  is validated by experiments given in
Section~\ref{sec:experiments}. Some non-crucial aspects are detailed in the
Appendix.\looseness-1
\subsubsection*{Related Work}

Skipping of complete subtrees has been considered before,
in several different contexts. For instance, the application
of the staircase join~\cite{DBLP:conf/vldb/GrustKT03} can be seen as
an instance  
of skipping: for the descendant axis, only the top-most independent
context nodes are considered, i.e., their subtrees are skipped;
in a similar way, even ancestor paths can be skipped by this join.
Skipping of subtrees is also common practice in advanced compilers
for pattern matching in programming languages. 
In \cite{lev03} selecting tree automata are compiled
into mutually recursive functions of an ML-style target language. 
They define ``loop breaker'' states, intuitively, a state
with transition $q,l\rightarrow(q,q)$. 
This is similar to non-relevant nodes, according
to our definition, and is used there to enforce the termination of the
generated code 
There is a large body of work on optimizations for evaluation of attribute
grammars (see, e.g., ~\cite{DBLP:journals/csur/Paakki95}) 
some of which correspond to skipping of subtrees; 
note that attribute grammars can simulate selecting TDTA and BDTAs.
In~\cite{Frischtcscompil} automata are used for tree pattern 
matching and subtrees are skipped according to 
type information.
Tree automata have been used for XPath, but mainly in the context
of streaming:
Koch~\cite{DBLP:conf/vldb/Koch03} runs BDTAs over a reversed XML
document followed  
by a top-down run, to evaluate XPath. Suciu \emph{et
al.}~\cite{DBLP:journals/tods/GreenGMOS04} use automata 
to evaluate many queries in parallel, over a stream.
We are not aware of any work that executes automata over
tree indexes, such as we do. In fact, even for usual DFAs over strings,
there is no prior work on executing DFAs or evaluating
regular expressions over indexed strings (where the index allows
to skip regions of the string, based on labels);
the closest work is~\cite{DBLP:journals/jacm/Baeza-YatesG96}.
Also comparable is the idea of running DFAs on grammar-compressed
strings. 
The THOR
system~\cite{DBLP:conf/sac/PettovelloF06,DBLP:conf/cascon/PettovelloF08}, 
uses data structures that support 
the same jumping operations as we do. However, they do step-wise evaluation of XPath a la
Koch and therefore cannot use these structures to restrict evaluation 
to only relevant nodes.\looseness-1

\label{sec:definitions}
\section{Selecting Tree Automata}
\label{sec:auto}
We define our notion of tree automata over binary trees.
When applying them to XML we use the well-known 
``first-child/next-sibling'' encoding: the first-child of a node
in the XML tree becomes the left child in the binary tree, and
the next-sibling of a node in the XML tree becomes the right
child in the binary tree. We also do not consider text nodes or
attributes (but a straightforward encoding is given in
\cite{DBLP:journals/corr/abs-0907-2089}). 
Let $\Sigma$ be an alphabet, i.e., a finite set of symbols.
The \emph{set of binary trees over $\Sigma$}, denoted $T(\Sigma)$,
is the smallest set $T$ such that (i) the leaf symbol $\#$ is in $T$ and
(ii) if $t_1,t_2\in T$ and $l\in\Sigma$, then 
$l(t_1,t_2)$ is in $T$. In the examples, we will often omit $\#$ for
concision.
A \emph{node} is a finite (possibly empty) sequence over $\{ 1,2\}$.
For a given tree $t\in T(\Sigma)$ its 
\emph{set of nodes}, denoted $\DOM{t}$, is the smallest finite set such that
(i) the empty sequence $\varepsilon$ is in $\DOM{t}$ and
(ii) if two sequences $\pi\cdot 1$ and $\pi\cdot 2$ are in $\DOM{t}$,
then $\pi\in \DOM{t}$. The label of the node $\pi$ in the tree $t$ 
is denoted by $t(\pi)$; 
for $t=l(t_1,t_2)$ it is defined as
$l$ if $\pi=\varepsilon$, and
as $t_i(\pi')$ if $\pi=i\cdot\pi'$; moreover,
for $t=\#$ we have $t(\varepsilon)=\#$.
As we can see, $\varepsilon$ denotes the root node,
and $\pi\cdot{}1$ and $\pi\cdot{}2$ denote the left and right-child
of the node $\pi$, respectively. When talking about the
\emph{followings} of a node $\pi$, we mean all the nodes visited after
$\pi$ during a pre-order traversal, that are not descendants of $\pi$.

\begin{definition}
A \emph{selecting tree automaton} (STA) $\mathcal{A}$ is a
6-tuple $\defaultAuto$ where
$\Sigma$ is an alphabet of \emph{input symbols},
$Q$ is a finite set of \emph{states},
$\mathcal{T}\subseteq Q$ is the set of \emph{top states},
$\mathcal{B}\subseteq Q$ is the set of \emph{bottom states},
$\mathcal{S}\subseteq Q\times\Sigma$ is the set of 
\emph{selecting configurations}, and $\delta$ is a finite
set of \emph{transitions}. A transition is tuple
$(q,L,q_1,q_2)$, where
$q,q_1,q_2\in Q$ and $L$ is a non-empty subset of $\Sigma$.\looseness-1
\end{definition}

From now on 
we let $\mathcal{A}=\defaultAuto$
 be a fixed (but arbitrary)
automaton, unless otherwise specified. 
We often write 
\mbox{$q,L\to(q_1,q_2)$} to denote that 
$(q,L,q_1,q_2)\in\delta$, and similarly
$q,L\Rightarrow(q_1,q_2)$ to denote that 
$(q,L,q_1,q_2)\in\delta$ and
$(q,l)\in\mathcal{S}$ for every $l\in L$.
Before defining the semantics of $\mathcal{A}$
via runs, we fix a few useful definitions.
Let $q,q_1,q_2\in Q$ and $l\in\Sigma$. The 
\emph{destination} and \emph{source states}, denoted
$\delta(q,l)$ and
$\delta(q_1,q_2,l)$, respectively, are defined as
$$
\begin{array}{l@{}c@{}l}
\delta(q,l)&=&\{ (q',q'')\mid\exists L\subseteq \Sigma \text{
  s.t. } l\in L \text{ and } (q,L,q',q'')\in\delta\}\\
\delta(q_1,q_2,l)&=&\{ q\mid\exists L\subseteq\Sigma \text{
  s.t. } l\in L \text{ and } (q,L,q_1,q_2) \in \delta\}.
\end{array}
$$
An automaton $\mathcal{A}$ is a 
\emph{top-down deterministic selecting tree automaton} (TDSTA) if
$\mathcal{T}$ is a singleton and,
for every $q\in Q$ and $l\in\Sigma$, 
$\delta(q,l)$ is a singleton.
Similarly, $\mathcal{A}$ is a 
\emph{bottom-up deterministic selecting tree automaton} (BDSTA) if
$\mathcal{B}$ is a singleton and,
for every $q_1,q_2\in Q$ and $l\in\Sigma$,
$\delta(q_1,q_2,l)$ is a singleton.
Note that if $\mathcal{S}$ is empty, then a TDSTA is exactly the same
as a classical deterministic top-down tree automaton (TDTA): the single
state in $\mathcal{T}$ is the initial state and the states in
$\mathcal{B}$ are the final states; similarly, a BDSTA is
a classical deterministic bottom-up tree automaton (BDTA): the single state
in $\mathcal{B}$ is its initial state and the states in
$\mathcal{T}$ are its final states.
The semantics of an STA is given by the set of trees it
recognizes (as for usual tree automata) and by the set of nodes it
selects. To formalize these notions, we introduce the concept of run.

\begin{definition}[Run of an STA]
Let $t\in T(\Sigma)$. 
A \emph{run} of $\mathcal{A}$ over $t$ is a total function
$R: \DOM{t}\rightarrow Q$
such that for all $\pi\in\DOM{t}$ with $t(\pi)\in\Sigma$,\\
\centerline{$R(\pi)\in\delta(R(\pi\cdot{}1),R(\pi\cdot{}2),t(\pi))$.}
The run $R$ is \emph{accepting} if and only if
\begin{myitemize}
\item $R(\varepsilon)\in\mathcal{T}$
\item for all $\pi\in\DOM{t}$ with $t(\pi)=\#$, 
$R(\pi)\in\mathcal{B}$. 
\end{myitemize}
We denote by $\Run{\mathcal{A}}{t}$ the set of all accepting
runs of $\mathcal{A}$ over $t$.
\end{definition}

An STA is \emph{top-down complete},
if for every $q\in Q$ and $l\in\Sigma$, 
$\delta(q,l)$ is non-empty. Similarly,
an STA is \emph{bottom-up complete},
if for every $q_1,q_2\in Q$ and $l\in\Sigma$,
$\delta(q_1,q_2,l)$ is non-empty. 
Top-down complete TDSTAs $\mathcal{A}$ and
bottom-up complete BDSTAs have a unique run for any input tree $t$. 

\begin{definition}
Let $\mathcal{A}$ be an STA. 
The \emph{language of $\mathcal{A}$}, denoted 
$\LANG{\mathcal{A}}$, is the set\\
\centerline{$
\LANG{\mathcal{A}} = \{t\in T(\Sigma)\mid
\Run{\mathcal{A}}{t}\neq\varnothing \}.
$}
The \emph{set of selected nodes of $\mathcal{A}$}, denoted
$\mathcal{A}(t)$, is the set\\[1mm]
\centerline{$
\mathcal{A}(t)=
\{\pi\in\DOM{t}\mid(R(\pi),t(\pi))\in\mathcal{S}\text{ and }
R\in\Run{\mathcal{A}}{t}\}.
$}
\end{definition}
We say that two STAs $\mathcal{A}$ and $\mathcal{A'}$ are
\emph{equivalent}, 
denoted $\mathcal{A}\equiv\mathcal{A'}$,
if $\LANG{\mathcal{A}} = \LANG{\mathcal{A}'}$ and
for every $t\in T(\Sigma)$, $\mathcal{A}(t)=\mathcal{A}'(t)$.

\begin{example}[STA for //{\tt a}//{\tt b}]
\label{ex:ssassb}
~\\
\centerline{$\mathcal{A}_{\text{//{\tt a}//{\tt
        b}}}=(
  \underset{\Sigma}{\{\mathtt{a},\mathtt{b},\mathtt{c}\}},
  \underset{Q}{\{q_0, q_1\}},
  \underset{\mathcal{T}}{\{q_0\}},
  \underset{\mathcal{B}}{\{q_0,q_1\}},
  \underset{\mathcal{S}}{\{(q_1,\mathtt{b})\}},
  \delta
)
$}\\[2mm]
\centerline{$\delta~=~$\small$
\begin{array}{l@{}c@{}l}
  q_0,\{\mathtt{a}\} & \rightarrow & (q_1,q_0) \\
  q_0,\Sigma\setminus\{\mathtt{a}\} & \rightarrow & (q_0,q_0)
\end{array}~~~~
\begin{array}{l@{}c@{}l}
q_1,\{\mathtt{b}\} & \Rightarrow & (q_1,q_1) \\
q_1,\Sigma\setminus\{\mathtt{b}\} & \rightarrow & (q_1,q_1) \\
\end{array}
$}
\end{example}
The TDSTA $\mathcal{A}_{\text{//{\tt a}//{\tt b}}}$ of 
Example~\ref{ex:ssassb} is 
\emph{not} deterministic bottom-up.
This is because its set $\mathcal{B}$ 
of bottom states is not a singleton.
In fact, we claim that there does not exist
any BDSTA that is equivalent to $\mathcal{A}_{\text{//{\tt a}//{\tt b}}}$, i.e.,
which selects the same nodes.
Intuitively, 
when a bottom-up automaton sees a \texttt{b}-node,
it does not know whether this node should be 
accepted or not (this depends on the existence of an
\texttt{a}-labeled ancestor).
We claim similarly that there exists BDSTAs for which
there is no equivalent TDSTA. The automaton implementing the query
//{\tt a}[.//{\tt b}] is such an example (which we detail in
Appendix~\ref{ap:auto}).
To conclude with the formal definitions, we characterize several kinds
of states that we use in the following sections.
\begin{definition}
\label{def:states}
Let $\mathcal{A}$ be an STA. 
A state $q\in Q$ is \emph{non-changing} if and only if
$\forall l\in\Sigma, \delta(q,l) = \{ (q,q)\}$. 
For a non-changing state $q$,
if $q\in\mathcal{B}$, $q$ is a \emph{top-down universal state};
if $q\in\mathcal{T}$, $q$ is a \emph{bottom-up universal state};
if $q\notin\mathcal{B}$, $q$ is a \emph{top-down sink state};
if $q\notin\mathcal{T}$, $q$ is a \emph{bottom-up sink state}.
\end{definition}

\noindent \textbf{Minimal Selecting Tree Automata} In
Appendix~\ref{ap:mini}
it is shown that for every TDSTA (resp. BDSTA)
there is a \emph{unique minimal} one, where minimal means with
the smallest number of states.
For a minimal TDSTA $\mathcal{A}$:
(i) at most one state is top-down universal state and
(ii) at most one state is a top-down sink state.
 If any of these states exist, then we denote them by
$q_\top$ and $q_\bot$, respectively.
The similar properties hold for BDTAs.
Another property that will be important for us in the next section
is that, in a minimal TDSTA or BDSTA, if 
a state $q$ is \emph{not} in $\{q_\top,q_\bot\}$, then
there must exist a label $l$ such that
$\delta(q,l)$ contains a pair different from $(q,q)$.
We say that $l$ is an \emph{essential label for $q$}
(in $\mathcal{A})$.


\section{Relevant nodes}
\label{sec:relnodes}

As we have explained in the Introduction, our goal is to improve query
answering time by reducing the number of nodes that have to be
visited by the evaluation function.
A common optimization technique for tree automata
(especially used in \emph{pattern-matching} and type-checking),
is to avoid visiting a subtree. 
For instance, consider the simple DTD
``\texttt{<!ELEMENT   a   ANY>}''
which states that an input document must have an \texttt{a}-labeled
root node and any well-formed content below it. A recognizer
automaton which checks the validity of a tree against this DTD is\\
\centerline{$
  \mathcal{A}=( \Sigma, 
  \underset{Q}{\{ q_0, q_\top, q_\bot\}},
  \underset{\mathcal{T}}{  \{  q_0 \}},
  \underset{\mathcal{B}}{\{ q_\top\}},\underset{\mathcal{S}}{\varnothing}, \delta)
$}\\[1mm]
\centerline{$\delta~=~$\small$
\begin{array}{l@{}c@{}l}
q_0,\{\texttt{a}\}&\rightarrow& (q_\top,q_\top)\\
q_0,\Sigma\setminus\{\texttt{a}\}&\rightarrow& (q_\bot,q_\bot)\\
\end{array}~~~~~
\begin{array}{l@{}c@{}l}
q_\top, \Sigma & \rightarrow & (q_\top,q_\top)\\
q_\bot, \Sigma & \rightarrow & (q_\bot,q_\bot)\\
\end{array}
$}
Since the automaton only changes state at the root node, 
only this node is ``relevant''; no information is gained at
any other node.
A clever evaluator may skip all non-relevant subtrees.
As we can see, whenever the automaton enters a \emph{non-changing state},
we can skip the current subtree.
Of course, there are automata equivalent to the one above which
change state in the subtrees under the root node (even though
this is not ``required''). How can we make sure that our automaton
only changes state when this is really necessary? 
The answer is simple: we \emph{minimize} the automaton.
If the minimal automaton changes state, then any other automaton
for the query does too; thus it uniquely determines the relevant nodes.
Moreover, as mentioned after Definition~\ref{def:states},
the minimal automaton has at most one state $q_\bot$ and one
state $q_\top$. It is therefore easy to determine when a
subtree can be skipped.
Of course, in a selecting tree automaton, all selected nodes must
be relevant, because we cannot select them without 
visiting them.
Consequently,
given a TDSTA ${\mathcal A}$ and a tree $t$ we say
that node $\pi$ of $t$ is relevant if the minimal automaton
${\mathcal A}_{\textit{min}}$ of $\mathcal{A}$ changes state at $\pi$.
We now give a general definition that can be used
for \emph{non-deterministic} automata; instead of minimality, the
definition uses equivalence between sub-automata.

\begin{definition}[Relevant nodes]
\label{def:noderel}
Let $\mathcal{A}$ be an STA. Let $t\in T(\Sigma)$ and
$R\in\Run{\mathcal{A}}{t}$. Let $\pi\in\DOM{t}$
such that $\pi\cdot 1\in \DOM{t}$ and $\pi\cdot 2\in
\DOM{t}$. 
The node $\pi$ is \emph{relevant} for the run
$R$ if and only
if either $(R(\pi),t(\pi))\in \mathcal{S}$ or \emph{none} of the following hold:
\begin{myitemize}
\item $\mathcal{A}[R(\pi)]\equiv\mathcal{A}[R(\pi\cdot
  1)]\equiv\mathcal{A}[R(\pi\cdot 2)]$;
\item $\mathcal{A}[R(\pi)]\equiv\mathcal{A}[R(\pi\cdot 1)]$ and $\mathcal{A}[R(\pi\cdot 2)]\equiv\mathcal{A}_\top$;
\item $\mathcal{A}[R(\pi)]\equiv\mathcal{A}[R(\pi\cdot 2)]$ and  $\mathcal{A}[R(\pi\cdot 1)]\equiv\mathcal{A}_\top$;
\end{myitemize}
where $\mathcal{A}_\top$ is such that
$\LANG{\mathcal{A}_\top}=T(\Sigma)$ and for all $t\in T(\Sigma)$,
$\mathcal{A}_\top(t)=\varnothing$. $\mathcal{A}[q]$ denotes the
restriction of $\mathcal{A}$ to $q$ (\emph{i.e.} where $\mathcal{T}$
is replaced by $\{ q \}$) and is formally defined in Appendix~\ref{ap:auto}.
\end{definition}
This definition generalizes the intuition we gave earlier. First, a selected node is
relevant. Then, a node can be skipped (\emph{i.e.} is \emph{not} relevant) if
the automaton performs the same computation on the node and on both
its children (informally the automaton ``loops'' both on the
left and right child). Or a node can be skipped if the automaton loops
on the left child and ``ignores'' the right child, \emph{i.e.} is in a
state that accepts $T(\Sigma)$ and does not mark any
node. Symmetrically, a node can be skipped if the automaton loops on
the right child and ignores the left one.
While Definition~\ref{def:noderel} gives a proper semantic
characterization of relevant nodes, we cannot use it to
derive an efficient evaluation procedure for STAs since:
\begin{myitemize}
\item[\emph{(i)}] it requires the accepting run to be known, while we
  want to deduce relevant nodes \emph{while} computing the run;
\item[\emph{(ii)}] it checks for equivalence of sub-STAs, an 
  EXPTIME-complete problem, even for recognizers.
\end{myitemize}
We present two exact algorithms for particular STAs, namely minimal TDSTAs and
minimal BDSTAs, and show how a particular index can be used to skip not only
subtrees but also internal nodes.\looseness-1

\subsection{Deterministic Top-Down Evaluation}
\label{sec:tdrel}
\subsubsection{Top-down Relevance}
As we have explained, testing the relevance of a node in the accepting
run of an automaton $\mathcal{A}$ consists in
checking the equivalence of several sub-automata.
It  is possible to perform this check efficiently for \emph{minimal}
TDSTAs. Indeed, in a minimal TDSTA, $q$ recognizes $T(\Sigma)$ if and only if $q$ is a
top-down universal state. More generally, given two states $q$ and $q'$
of $\mathcal{A}$:\\
\centerline{$\mathcal{A}[q]\not\equiv\mathcal{A}[q'] \Longleftrightarrow q\neq q'.$}
This is a consequence of the definition of a minimal automaton. Given
a TDSTA and a run, we can easily characterize the set of relevant nodes:
\begin{lemma}[Top-down relevant nodes]
\label{lem:tdrel}
Let $\mathcal{A}$ be a
minimal top-down complete TDSTA, $t\in T(\Sigma)$,  $R\in \Run{\mathcal{A}}{t}$ and
 $\pi\in\DOM{t}$ such that $\pi\cdot 1\in \DOM{t}$ and $\pi\cdot
  2\in\DOM{t}$. $\pi$ is \emph{top-down relevant} in $R$ if and only
  if either $(R(\pi),t(\pi)) \in \mathcal{S}$ or if none of the following hold:
\begin{myitemize}
\item $R(\pi) = R(\pi\cdot 1) = R(\pi\cdot 2)$
\item $R(\pi) = R(\pi\cdot 1)$ and $R(\pi\cdot 2)=q_\top$
\item $R(\pi) = R(\pi\cdot 2)$ and $R(\pi\cdot 1)=q_\top$
\end{myitemize}
\end{lemma}
For a given run of a minimal TDSTA, the relevant nodes
are either the selected nodes or nodes for which a state-change
occurs. An important observation is that for TDSTAs, a state
change is exactly determined by the set of \emph{essential}
labels. For instance, in the automaton $\mathcal{A}_{\text{//{\tt
      a}//{\tt b}}}$ of Example~\ref{ex:ssassb}, the set of essential labels
for state $q_0$ is $\{ \texttt{a} \}$: the automaton changes
state only if it encounters an \texttt{a}-labeled node during the
top-down run. 
\subsubsection{Top-Down Jumping Functions}
\label{sec:tdjump}
Based on this observation, we define particular jumping
functions in a tree which extend the basic firstChild and
nextSibling moves.
The implementation of
 such functions using state of the art tree indexes is later discussed
 in Section~\ref{sec:experiments}.

\begin{definition}[Top-down jumping functions]
  Let $t$ be a tree in $T(\Sigma)$. We define the
    functions $\mathbf{d}_t$, $\mathbf{f}_t$, $\mathbf{l}_t$, $\mathbf{r}_t$ as:
  \begin{myitemize}
    \item $\mathbf{d}_t : \DOM{t}\times 2^\Sigma \rightarrow
      \DOM{t}\cup\{\Omega\}$ where $\mathbf{d}_t(\pi,L)$ returns the first descendant $\pi'$ of $\pi$ (in
      document-order) such that $t(\pi')\in L$;

    \item $\mathbf{f}_t : \DOM{t}\times 2^\Sigma\times \DOM{t}
      \rightarrow  \DOM{t}\cup\{\Omega\}$ where
      $\mathbf{f}_t(\pi,L,\pi_0)$ returns the first following node
      $\pi'$ of $\pi$ such that $\pi'\in L$ and $\pi'$ is a descendant
      of $\pi_0$.

    \item $\mathbf{l}_t : \DOM{t}\times 2^\Sigma \rightarrow
      \DOM{t}\cup\{\Omega\}$ where
      $\mathbf{l}_t(\pi,L)$ returns the first descendant $\pi'$ of
      $\pi$ whose label is in $L$ and such that $\pi'=\pi\cdot
      1\ldots\cdot 1$ (left-most path);

    \item $\mathbf{r}_t : \DOM{t}\times 2^\Sigma \rightarrow
      \DOM{t}\cup\{\Omega\}$ where
      $\mathbf{r}_t(\pi,L)$ returns the first descendant $\pi'$ of
      $\pi$ whose label is in $L$ and such that $\pi'=\pi\cdot
      2\ldots\cdot 2$ (right-most path).

  \end{myitemize}
All these function returns a special error node $\Omega$ if there is
no $\pi'\in \DOM{t}$ which fits their definitions.
\end{definition}
Using these functions, the set of top-most nodes ${\pi_0,\ldots,\pi_n}$
whose labels are in $L$,  in a subtree rooted at $\pi$ can be computed by:\\
\centerline{$\pi_0=\mathbf{d}_t(\pi,L)$ and then $\pi_{n+1}
=\mathbf{f}_t(\pi_{n},L,\pi)$, until $\pi_n=\Omega$.}

\subsubsection{Jumping Top-Down Algorithm}
We use the jumping functions defined in the previous section to compute a partial
run for a minimal TDSTA and an input tree $t$. More
specifically, the algorithm returns a mapping from nodes to states. If
there is
no accepting run, the algorithm aborts and returns an empty mapping.
We describe informally the algorithm (its pseudo code is given in Appendix~\ref{alg:tdj}).
The algorithm is implemented by the mean of a recursive function
\textit{topdown\_jump}
which takes as argument a node $\pi$ in the input tree $t$ and a state
$q$
(initially the root node $\varepsilon$ and the initial state $q_0$ of the
TDSTA). This function works like the usual top-down evaluation
procedure for a TDSTA. First, if $\pi$ is a leaf (a $\#$-labeled node
in our context) then  the automaton checks whether
$q\in\mathcal{B}$. If this is the case, the function returns the
mapping $\{\pi\mapsto q\}$ and fails otherwise. More interestingly if
$\pi$ is not a leaf, then function computes the states
$(q_1,q_2)=\delta(q,t(\pi))$. If either $q_1$ or $q_2$ is the sink state,
then the function fails (there is no accepting run).
Otherwise, the function performs a case analysis on $q_i$ to determine
the set of top-most relevant nodes in the subtree rooted at $\pi\cdot
i$ (for $i\in\{ 1,2\}$). The function considers the three cases given
in Lemma~\ref{lem:tdrel}:
\begin{myitemize}
  \item $q_i,L'\rightarrow (q_i,q_i)$ and $q_i,L\rightarrow (q',q'')$
    with $q'$ or $q''$ distinct from $q_i$. The function performs its
    recursion on all the top-most descendants of $\pi\cdot i$ whose
    label is in $L$;
\item $q_i,L'\rightarrow (q_i,q_\top)$ and $q_i,L\rightarrow (q',q'')$
  with $q'$ distinct from $q_i$. The function is called recursively 
  on the node $\mathbf{l}_t(\pi\cdot i,L)$ (the automaton loops on the
  left-most path below the current node).
\item $q_i,L'\rightarrow (q_\top,q_i)$
  and $q_i,L\rightarrow (q',q'')$ and $q''$ distinct from $q_i$.
  The function is called recursively  on the node $\mathbf{r}_t(\pi\cdot i,L)$
\end{myitemize}
If none of the above hold, $\pi\cdot i$ is relevant and the function
is recursively called on $\pi\cdot i$ itself.
Lastly, the function returns the mapping $\{ \pi \mapsto q\}$
augmented by the mappings returned by the recursive calls on the left and
right subtrees.
This function computes the optimal traversal
with respect to relevant nodes:
\begin{theorem}
  \label{thm:tdexact}
  Let $t\in T(\Sigma)$. Let
  $\mathcal{A}$
  be a minimal TDSTA. Let $R$ be the run of $\mathcal{A}$ over
  $t$ and $R' = \textit{topdown\_jump}(t,\mathcal{A})$. 
  \begin{myitemize}
  \item if $R$ is an accepting run, then for all $\pi\in\DOM{t}$,
    $R'(\pi)=R(\pi)$ if an only if $\pi$ is top-down relevant for $R$;
  \item if $R$ is not an accepting run, then $R'=\varnothing$.
    \end{myitemize}
\end{theorem}

\subsection{Deterministic Bottom-Up Evaluation}
\label{sec:bujump}
While a top-down run of an automaton can be
translated into a natural top-down tree traversal, bottom-up runs are
more complicated.
Assuming that a parent
move and access to the sequence of leaves of an input tree are
supported, we can devise a ``pure bottom-up'' evaluation function,
which starts from the sequence of leaves and works its way up to the
root. The pseudo code of this algorithm is given in
Appendix~\ref{ap:bu}. 
From the sequence
$(\pi_1,q_0),\ldots,(\pi_n,q_0)$ of all leaves $\pi_i$ and initial
state $q_0$ the algorithm proceeds to ``reduce'' them (by
replacing two siblings by their parent and corresponding state) until
the root node is
obtained. If the first two nodes in the current list
are not siblings, the algorithm first reduces recursively the tail of
the list, pushes back the first element on the reduced tail (whose
size decreased) and reduces the new list.
For BDSTA, relevance is once again defined in terms of state change,
but in a more complex way.
\begin{lemma}[Bottom-up relevant nodes]
\label{def:burelnodes}
Let $\mathcal{A}$ be a bottom-up complete
minimal BDSTA. Let $\mathcal{B}=\{ q_0\}$. Let $t$ be a tree. Let
$R$ be the
accepting run for $\mathcal{A}$ and $t$ (if it exists). Let
$\pi\in\DOM{t}$ such that $\pi\cdot 1\in\DOM{t}$ and $\pi\cdot 2\in\DOM{t}$.
The node $\pi$ is \emph{relevant} if and only if
$(R(\pi),t(\pi))\in\mathcal{S}$ or none the following conditions holds:
\begin{myitemize}
\item $R(\pi) = q_\top$
\item $R(\pi)=R(\pi\cdot 1)=R(\pi\cdot2)$;
\item $R(\pi)=R(\pi\cdot 1)$ and $R(\pi\cdot 2)\in\{q_0,q_\top\}$;
\item $R(\pi)=R(\pi\cdot 2)$ and $R(\pi\cdot 1)\in\{q_0,q_\top\}$;
\end{myitemize}
\end{lemma}
We do not give the proof that these conditions on states coincide with the relevance
of nodes as given by Definition~\ref{def:noderel}, but illustrate them
by an example given in Appendix~\ref{ap:bu}.

In the same way  we generalized firstChild to $\mathbf{d}_t$
and $\mathbf{l}_t$ and
nextSibling to $\mathbf{f}_t$ and $\mathbf{r}_t$ for the
 top-down case, the
 moves used in the bottom-up algorithm can be generalized. The sequence of all leaves is replaced by the
sequence of bottom-most nodes with a particular label and the parent move
can be replaced by either a jump to an ancestor with a particular
label, or the restriction of this jump to the left-most or right-most
path leading to the current node. Also, testing whether two nodes are
siblings in generalized into getting the common ancestor of two
nodes. We dub the generalized bottom-up jumping algorithm
\textit{bottomup\_jump}, but the many cases it handles (intuitively,
when trying to jump above two nodes $\pi_1$ and $\pi_2$ we must not
jump above their common ancestor, or we could miss some nodes) makes
its presentation verbose even in the form of pseudo-code. Second, the
tree indexes that we use in our implementation do not implement the
\emph{ancestor} jumps efficiently (they amount to a sequence of parent
calls). We therefore limit ourselves to state the existence of
algorithm~\textit{bottomup\_jump}, and give its theoretical
properties:
\begin{theorem}
  \label{thm:buexact}
  Let $t\in T(\Sigma)$. Let
  $\mathcal{A}$
  be a minimal BDSTA. Let $R$ be the run of $\mathcal{A}$ over
  $t$ and $R' = \textit{bottomup\_jump}(t,\mathcal{A})$. 
  \begin{myitemize}
  \item if $R$ is an accepting run, then for all $\pi\in\DOM{t}$,
    $R'(\pi)=R(\pi)$ if an only if $\pi$ is bottom-up relevant for $R$;
  \item if $R$ is not an accepting run, then $R'=\varnothing$.
    \end{myitemize}
\end{theorem}

\section{Automata for XPath}
\label{sec:implem}
We present in this section our compilation target for XPath
expressions, namely alternating selecting tree automata (ASTA).
We then consider a particular fragment of XPath for which we illustrate our
compilation scheme. Afterwards we introduce a technique for evaluating
an ASTA in a jumping fashion, using a sound approximation of the sets of
relevant nodes of the query. We also present various implementation
techniques to further improve the complexity in practice of the
evaluation of ASTAs.

\subsection{Alternating Selecting Tree Automata}
We introduce a compact variation of STAs which
works with Boolean formulas over states. 
\begin{definition}[Alternating Selecting Tree Automata (ASTA)]~~\\
An ASTA $\mathcal{A}$ is a tuple
$(\Sigma,\mathcal{Q},\mathcal{T},\delta)$, where
 $\Sigma$ is the alphabet of input symbols,
 $Q$ is the finite set of states,
 $\mathcal{T}\subseteq Q$ is the set top states, and
 $\delta$ is a set of tuples $(q,L,\tau,\phi)$, called transitions,
 where $q\in Q$, $L\subseteq \Sigma$, $\tau\in
 \{\rightarrow,\Rightarrow\}$ and $\phi$ is a \emph{Boolean formula} 
 generated by the following EBNF.\\
\centerline{
$\begin{array}{lcll}
    \phi & ::= & \top ~|~ \bot ~|~ \phi\lor\phi ~|~ \phi\land\phi ~|~
    \lnot \phi ~|~ \downarrow_1 q ~|~ \downarrow_2 q & (q\in Q)\\
  \end{array}$
}
\end{definition}
The semantics of such automata combine the rules for a
classical alternating automaton, with the rules of a selecting tree
automaton. The complete rules for the evaluation of formula and the
selection of nodes is given in Appendix~\ref{ap:implem}.

\subsection{From XPath to Automata}
The fragment of XPath we consider in this presentation is the forward
fragment of Core XPath, containing \texttt{descendant} and
\texttt{child} axes as well as arbitrarily nested predicates using
\texttt{or}, \texttt{and} and \texttt{not} Boolean connective over
path expressions. The full EBNF description of this fragment is given
in Appendix~\ref{ap:implem}.
We illustrate how to compile an XPath expression of this fragment into an ASTA.
\begin{example}[{ASTA for the query //{\tt a}//{\tt b}[{\tt c}]}]~Let\\
\label{ex:xpathcomp}
\centerline{$\mathcal{A}_{\text{//{\tt a}//{\tt b}[ {\tt c} ]}}=(\Sigma,\{q_0,q_1,q_2\}, \{q_0\},\delta)$}
where $\delta$ is:\\
\centerline{\small$
\begin{array}{l@{}c@{}l}
  q_0, \{ a\} & \rightarrow & \downarrow_1 q_1\\
  q_0, \Sigma&\rightarrow &
  \downarrow_1 q_0 \lor \downarrow_2 q_0\\
\end{array}\rule[-10pt]{0.5pt}{20pt}
\begin{array}{l@{}c@{}l}
  q_1, \{ b\} & \Rightarrow & \downarrow_1 q_2\\
  q_1, \Sigma&\rightarrow &
  \downarrow_1 q_1 \lor \downarrow_2 q_1\\
\end{array}\rule[-10pt]{0.5pt}{20pt}
\begin{array}{l@{}c@{}l}
  q_2, \{ c\} &\rightarrow & \top\\
  q_2, \Sigma &\rightarrow & \downarrow_2 q_2\\
\end{array}
$}
\end{example}
It is easy to see with this example that such automata can be built by
a simple traversal of the parse tree of the XPath query.
The compilation scheme we follow associates one state for each
step of the query, and each state has at most two transitions. The first
one represents a ``progress'' from the current step to the next
step (in the XPath query). The second transition represents a
recursion on the first child,
the second child or both.
Note that non-determinism is used here in an essential way.
 For instance, in $\mathcal{A}_{\text{//{\tt a}//{\tt b}[{\tt c}]}}$, in
 state $q_1$, if the current node is labelled {\tt b}, then
the automaton selects a node if its first child is in state $q_2$
and at the same time remains in state $q_1$ for both the first child and
the second child. 

While this automaton does not seem to justify the use of alternation,
we give in Appendix~\ref{ap:implem} a query whose corresponding ASTA
is linear in size but
whose STA (even non-deterministic) is exponentially larger.

On this example, we observe that the particular ASTAs we consider
share many common traits with the minimal deterministic TDSTAs of
Section~\ref{sec:tdrel}. First a state change occurs whenever the
automaton gains new knowledge toward answering the query. Second, a
top-down universal state correspond to the presence of $\top$ in a
formula (that is, $(q_\top,q_\top)$) or the absence of a
$\downarrow_1$ or $\downarrow_2$ move (for instance $\downarrow_2 q$
is the counterpart of $(q_\top, q)$ in our previous model). In such
automata, a state change has the same meaning as in a minimal
deterministic one.

\subsection{Bottom-Up Evaluation with Top-Down Pre-Processing and
  Jumping}
\label{ssec:butd}
Before discussing how to evaluate such automata using only relevant
nodes, we give a ``non-jumping'' run function for ASTAs.
\lstset{language=[Objective]Caml,
  basicstyle=\small,
  columns=flexible,
  identifierstyle=\it,
  keywordstyle=\ttfamily,
  morekeywords={case,of,end,return,left,right,is_root,hd,tl,parent},
  numbers=left,
  numberstyle=\scriptsize,
  mathescape=true,
  xleftmargin=0pt,
}
\begin{algorithm}[Evaluation of an ASTA]~\\[1mm]
  \begin{minipage}{9cm}
    \small
    \label{alg:topdown}
    \textbf{Input:} $\mathcal{A}=(\Sigma,Q,\mathcal{T},\delta)$, $t$, $\pi$, $r$\qquad\textbf{Output:} $\Gamma$\\
    where $\mathcal{A}$ is the automaton, $t$ the input tree, $r$ a set
    of states and\\ $\Gamma$ is a result set. Initially $\pi = \varepsilon$
    and $r=\mathcal{T}$.
\end{minipage}
\begin{minipage}{\textwidth}
\begin{lstlisting}
         function eval_asta $(\mathcal{A},t,\pi,r)$ =
            if $t(\pi) = \#$ then return $\emptyset$ else
            let trans = $\{ (q,L,\tau,\phi)\in\delta\mid q\in r \textrm{~and~} t(\pi) \in L\}$ in
            let $r_i$ = $\{ q\mid\downarrow_i q \in \phi, \forall \phi \in \textit{trans}\}$ in
            let $\Gamma_1$ = eval_asta $(\mathcal{A},t,\pi\cdot 1,r_1)$ 
            and $\Gamma_2$ = eval_asta $(\mathcal{A},t,\pi\cdot 2,r_2)$ 
            in return $\texttt{eval\_trans}(\Gamma_1,\Gamma_2,\pi,\textit{trans})$
\end{lstlisting}
\end{minipage}
\end{algorithm}
The function \textit{eval\_asta} evaluates an ASTA over an input tree
$t$. It returns a result set $\Gamma$ which is a mapping from states
to the sets nodes selected in that state. In the usual non-selecting,
algorithm, $\Gamma$ is
simply the set of states which accept the current node $\pi$. 

We have already described in details how node selection works for such
automata in \cite{DBLP:journals/corr/abs-0907-2089}, we focus on the
main novelty of this work, relevant node approximation. The interested
reader can refer to Appendix~\ref{ap:implem} for the complete semantics of ASTA
(including node selection) as well as a commented example. This process is
abstracted by the function \textit{eval\_trans} on Line~7 which
handles both selection and evaluation of formulas.

\begin{figure}
\centerline{\includegraphics[width=6.6cm]{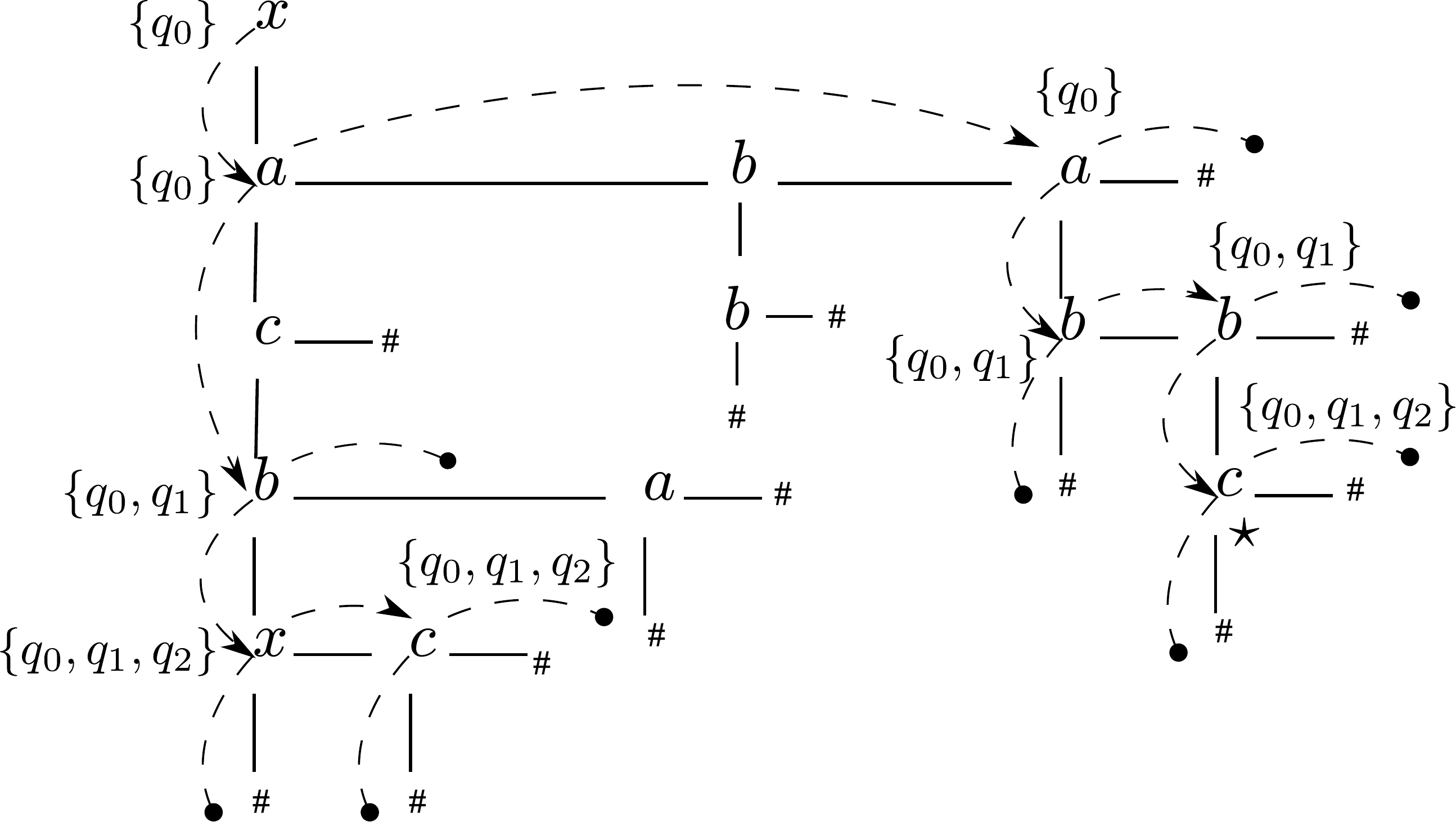}}
\begin{scriptsize}
$\begin{array}{l@{\rightarrow}l}
\{ q_0\}, \{ a \} & \{ q_0,q_1\}, \{q_0\}\\
\{ q_0\}, \Sigma\setminus\{a\} & \{ q_0\}, \{q_0\}\\
\{ q_0,q_1\}, \{ b \} & \{ q_0,q_1,q_2\}, \{q_0,q_1\}\\
\{ q_0,q_1\}, \Sigma\setminus\{b\} & \{ q_0,q_1\}, \{q_0,q_1\}\\
\{ q_0,q_1,q_2\}, \{ b \} & \{ q_0,q_1,q_2\}, \{q_0,q_1,q_2\}\\
\{ q_0,q_1,q_2\}, \{ c \} & \{ q_0,q_1\}, \{q_0,q_1\}\\
\{ q_0,q_1,q_2\}, \Sigma\setminus\{b\} & \{ q_0,q_1\}, \{q_0,q_1,q_2\}\\
\end{array}$
\end{scriptsize}
\caption{Top-down approximation for \texttt{//a//b[c]} and
  corresponding jumps}
\label{fig:tdaex}
\vspace{-5mm}
\end{figure}

The parameter $r$ of the function 
\textit{eval\_asta} allows one to restrict bottom-up runs of $\mathcal{A}$
to only those which end-up in a top-state at the root node.
What this algorithm does is to run first a \emph{deterministic top-down}
automaton $\mathcal{A}_{\text{approx}}$ during the recursive
descent. This automaton is a sound
approximation of $\mathcal{A}$ in the sense that for any $t\in
T(\Sigma)$, $t\notin\LANG{\mathcal{A}_{\text{approx}}} \Rightarrow
t\notin \LANG{\mathcal{A}}$. We can make further use of this automaton
$\mathcal{A}_{\text{approx}}$ by only jumping to a super-set of its
relevant nodes.
\begin{definition}[Top-down approximation]
Let $\mathcal{A}=(\Sigma,Q,\mathcal{T},\delta)$ be an ASTA. The
top-down approximation of $\mathcal{A}$ is the automaton
$\textit{tda}(\mathcal{A})=(\Sigma, 2^{Q}, \{ \mathcal{T} \},
\delta_a)$ where\\[1mm]
\centerline{$
\begin{array}{l}
\delta_a = \{ (S,{\sigma},\rightarrow,S_1,S_2)\mid S\subseteq Q, \sigma\in\Sigma,\hspace{3cm}\\
\multicolumn{1}{r}{S_i = \{ q\in Q\mid \exists q'\in S, \downarrow_i q
  \in\delta(q',\sigma)\}\}}\\
\end{array}
$}
\end{definition}
The exponential blow-up exhibited by this construction is avoided
by computing the top-down approximation on-the-fly. 
 The interesting part is now: what relevant nodes can be computed ---and
therefore which jumps can be performed--- if we consider the states in
$\textit{tda}(\mathcal{A})$. Figure~\ref{fig:tdaex} illustrates the
top-down approximation for the automaton
$\mathcal{A}_{\text{//{\tt a}//{\tt b}[{\tt c}]}}$  as well as the jumps that can be
computed from its non-changing states.
As we can see in the figure, the top-down approximation
allows us to jump quite precisely in the tree. If the destination
state for a subtree is $\{q_0\}$ the automaton can jump to the top-most
$a$ node in the subtree. If the destination state is $\{q_0,q_1\}$,
the automaton can jump to a top-most {\tt b} node in the
subtree. If the destination state is $\{ q_0,q_1,q_2\}$, no jump is
possible, the automaton must perform a firstChild or nextSibling
move. However, once in state $\{ q_0,q_1,q_2\}$, if the label is {\tt c}
then the automaton returns in state $\{q_0,q_1\}$ and can therefore
jump to find new {\tt b} nodes.
\subsection{Implementation Techniques}

\noindent \textbf{Hybrid Evaluation} The main drawback of the
top-down approximation of relevant nodes is to force a ``top-down
view'' of the query. For instance for query //{\tt a}//{\tt b}[{\tt
  c}],
if a document contains a lot of {\tt a}-nodes and few {\tt b} nodes,
the former ones will be needlessly visited since they are part of the
top-down approximation of the relevant nodes.
 To alleviate this problem, we
propose an alternative evaluation strategy dubbed 
hybrid evaluation. The idea is to start anywhere
in the query and the document. In the case of query //{\tt a}//{\tt b}[{\tt c}],
this means starting evaluation at all \texttt{b}-nodes in the
document, and check in a recursive top-down+bottom-up fashion the 
filter ``[\texttt{c}]'' in their subtrees  and the
path ``//\texttt{a}'' in their upward context. Such strategy can be 
effective if the count of {\tt b}-nodes is low.

\noindent\textbf{Memoization} If we consider
Algorithm~\ref{alg:topdown}, we see that the
computations performed at Line~3 (and~7) have complexity
$O(|\delta|)$. They contribute the $|Q|$ factor to the complexity
$O(|Q|\cdot|D|)$ of the evaluation function. We can memoize
these computations which only depends on $r$ and $t(\pi)$ for Line~3
and $r$,$t(\pi)$, $r_1$ and $r_2$ for Line~7.
This technique amortises the $|Q|$ factor over the whole run:
except for a few ``warm-up'' nodes for which the
all the transitions must be scanned, the rest of the run consists of a
succession of look-ups in a table, one for each node visited during
the run.


\noindent\textbf{Information Propagation} During the traversal, a
node is ``seen'' three times by the
evaluation function: $(i)$ when reaching the node during the top-down
traversal, $(ii)$ when returning from the evaluation of the first
child $(iii)$ when returning from the evaluation of the second
child. Instead of waiting $(iii)$ to evaluate the
transitions, we can already evaluate them in $(ii)$ having only the
knownledge for the first child. This reduces the number of states to
verify while visiting the second child. In particular it ensures that
for an XPath predicate, only one witness is checked by the automaton,
the first one in pre-order (existential semantics).
This
is inspired from the evaluation of Non-Uniform Automata of
\cite{Frischtcscompil}.

\noindent\textbf{Result Sets} Since the nodes are traversed in
document order and only once, result sets can be implemented as simple
lists with constant time concatenation for the union of two
result-sets. 

\begin{figure}
\begin{scriptsize}
\begin{tabular}{l@{~}l}
Q01 & \textrm{/site/regions}\\
Q02 & \textrm{/site/regions/europe/item/mailbox/mail/text/keyword}\\
Q03 &
\textrm{/site/closed\_auctions/closed\_auction/annotation/description/parlist/listitem}\\
Q04 & \textrm{/site/regions/*/item}\\
Q05 & \textrm{//listitem//keyword}\\
Q06 & \textrm{/site/regions/*/item//keyword}\\
Q07 & \textrm{/site/people/person[ address and (phone or homepage) ]}\\
Q08 & \textrm{//listitem[ .//keyword and .//emph]//parlist}\\
Q09 & \textrm{/site/regions/*/item[ mailbox/mail/date ]/mailbox/mail}\\
Q10 & \textrm{/site[ .//keyword]}\\
Q11 & \textrm{/site//keyword }\\
Q12 & \textrm{/site[ .//keyword ]//keyword}\\
Q13 & \textrm{/site[ .//keyword or .//keyword/emph ]//keyword}\\
Q14 & \textrm{/site[ .//keyword//emph ]/descendant::keyword }\\
Q15 & \textrm{/site[ .//*//* ]//keyword}\\
\end{tabular}
\end{scriptsize}
\caption{Tree queries used in the experiments}
\label{fig:queries}
\vspace{-3mm}
\end{figure}

\section{Experiments}
\label{sec:experiments}

We use several experiments to illustrate the behaviour of the
algorithms we introduced and gauge precisely the impact of each of the
optimizations and implementation techniques we presented.
Due to space constraints, we do not try to give in this paper the bare
performances of our implementation. The interested reader can refer to 
\cite{DBLP:journals/corr/abs-0907-2089} where a large experimental
section compares our implementation to state of the art query engines
(MonetDB/XQuery and Qizx/DB), for a richer set of queries (both tree
oriented and text oriented). Nevertheless, we provide for the
sake of completeness a comparison of our implementation with the
MonetDB/XQuery engine in Appendix~\ref{ap:experiments}.

\begin{figure*}[tbph]
\begin{small}
\hspace{2cm}\begin{tabular}{|l@{~}*{15}{|@{~}c@{~}}|}
\hline
	&Q01	&Q02	&Q03	&Q04	&Q05	&Q06	&Q07	&Q08
        &Q09	&Q10	&Q11	&Q12	&Q13	&Q14	&Q15\\
\hline
\textbf{(1)}	&1	&3518	&8860	&22620	&36511	&42955	&9885	&5026	&21851	&1	&73070	&73070	&73070	&73070	&73070\\
\textbf{(2)}	&2	&27943	&42333	&22628	&76391	&65583	&66256	&75727	&80846	&2	&73071	&73071	&73071	&73072	&73074\\
\textbf{(3)}	&20	&353122	&422060	&67898	&\# nodes	&1892764	&515305	&\# nodes	&1030955	&33	&\# nodes	&\# nodes	&\# nodes	&\# nodes	&\# nodes\\
\textbf{(4)}	&4	&24	&20	&19	&7	&24	&33	&20	&32	&4	&5	&7	&7	&11	&9\\
\textbf{(5)}	&50	&12.5	&20.9
&99.9	&47.7	&65.4	&14.9
&6.63	&27.0	&50	&99.9	&99.9
&99.9	&99.9	&99.9\\
\hline
\end{tabular}\\
\textbf{(1)}: Number of selected nodes \hfill
\textbf{(2)}: Number of visited nodes with jumping\hfill
\textbf{(3)}: Number of visited nodes without jumping\\
\textbf{(4)}: Number of memoized transitions\hfill
\textbf{(5)}: Ratio of selected nodes vs. approximated top-down
relevant nodes (in \%)\hfill\# nodes = 5673051\\
\end{small}
\vspace{-7mm}
\caption{Number of selected and visited nodes (w and wo jumping), and
  number of memoized configurations}
\label{fig:numbers}
\end{figure*}
\noindent \textbf{Implementation}
 Our implementation\footnote{Our implementation is written in OCaml (for ASTA/XPath query part)
and C++ (for the indexes). Our test machine is described in Appendix~\ref{ap:experiments}.
}
features a bottom-up with top-down pre-processing evaluation function
(``top-down+bottom-up'' as we refer to it in the rest of the section) which uses the jumping
primitives described in
\cite{DBLP:journals/corr/abs-0907-2089}. These indexes support
jumping to the first descendant and following nodes whose label is in a
set $L$ in time $O(|L|)$. As for the hybrid evaluation function, due to
the lack of upward-jumping functions in this index, it performs its
upward part using only parent
moves (instead of jumping to ancestors with particular labels).
It however remains an effective strategy when one of the labels in the
query has a low count (our index provides the global count of a label
in constant time).


\noindent\textbf{Documents and Queries} We used the XMark \cite{xmark},
document generator for our tests. We report our results for a document of
size 116MB. The tree oriented queries we used are given in
Figure~\ref{fig:queries}.
Q01 to Q09 are realistic queries for XMark documents, taken from the
XPathMark benchmark \cite{xpathmark}. Q10 to Q15 allow
us to illustrate in more details the behaviour of our ASTAs.

\begin{figure}
\vspace{-5mm}
\includegraphics[width=\columnwidth]{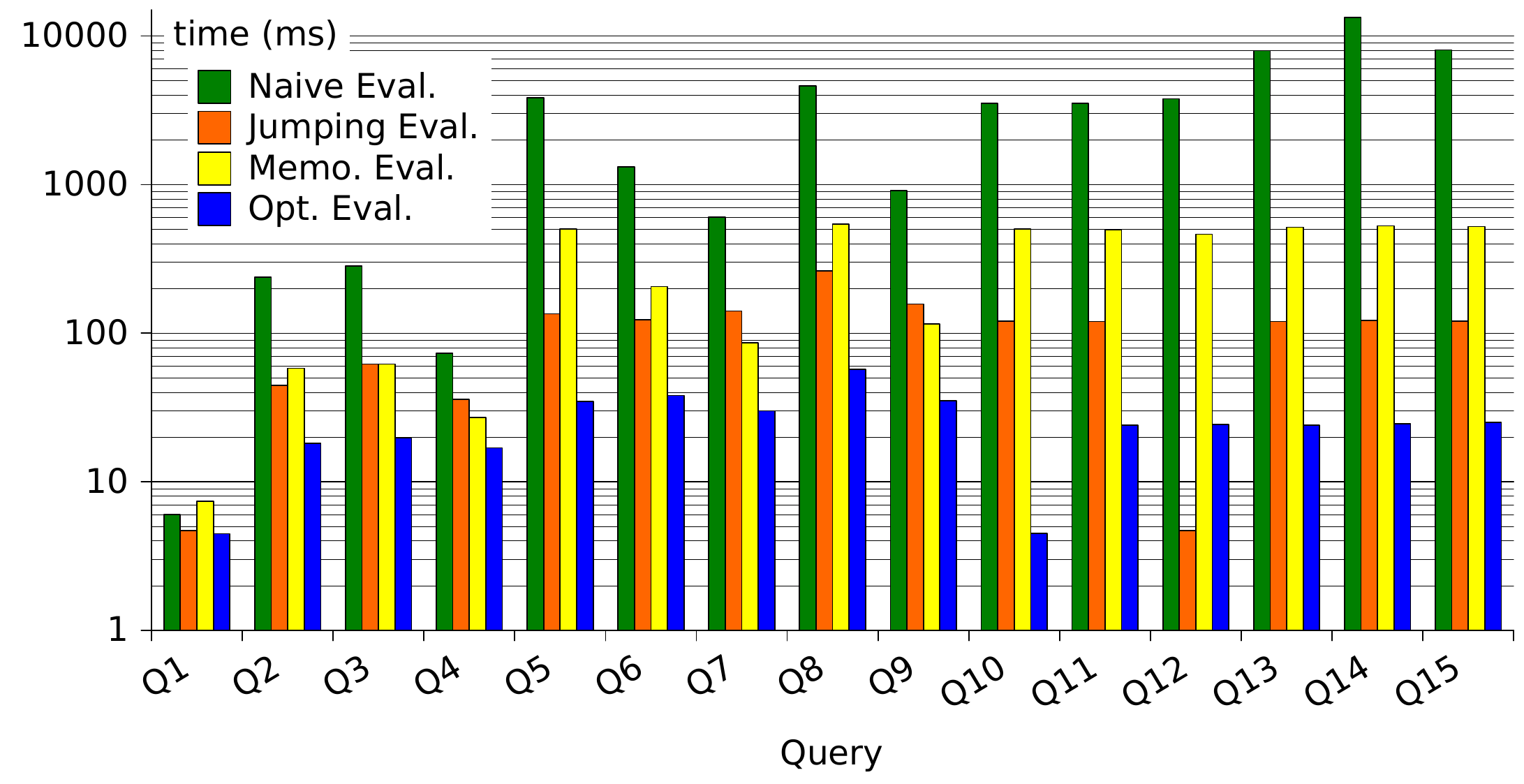}
\vspace{-5mm}
\caption{Impact of the jumping and memoization on query evaluation time}
\label{fig:feat}
\end{figure}

\noindent\textbf{Impact of Jumping and Memoization} We report in
Figure~\ref{fig:feat} the query answering time of our
engine for each query (note the logarithmic scale for the times).
 The``Naive Eval.'' series represents a straightforward execution of
Algorithm~\ref{alg:topdown}. As we can see,
a naive evaluation where the $|Q|$ factor has to be paid for each
node, and which potentially visits every node  in $D$ is not
satisfactory. For queries where a ``//'' occurs at top-level,
the full document needs to be traversed, yielding an evaluation time from
1s to 10s.
The ``Jumping
Eval.'' series represents a run where the evaluation function computes
the top-down approximation of relevant nodes on-the-fly and jumps only
to these
nodes. No memoization occurs therefore the $|Q|$ factor is paid for
each visited node. As expected, this is a huge improvement compared to
the naive case. With this optimization alone, all the tested
queries require less than 150ms to evaluate, an improvement of ten to
hundred-folds.
The ``Memo. Eval.'' series represents runs
where on-the-fly computations are memoized. For these runs, the
$|D|$ factor is paid in full (unless the automaton can skip whole
subtrees as in Q01) while the $|Q|$ factor is amortized.
This technique also improve query answering time considerably: a full
traversal takes no more than 450ms. The
fact that only firstChild and nextSibling moves are used also
demonstrate that alternating automata are a framework of choice, even
over pointer-based data-structures.
Lastly, the ``Opt. Eval'' series represents runs where both
optimizations are enabled. We can see that they are complementary:
with the exception of Q01 and Q12, the ``Opt. Eval'' time is always
better (at least twice as fast) as either optimization taken
individually. Q01 and Q12
are a very particular case where the query only touches two nodes
therefore the transitions memoized in the look-up table are never
re-used and their insertions only constitute an overhead.

\noindent\textbf{Top-Down Relevance Approximation, Automata Logic and
Memoization}: the table in Figure~\ref{fig:numbers}
gives the number of selected nodes (Line~\textbf{(1)}).
These numbers are to be contrasted with Line~\textbf{(2)}, which
represents the number of nodes visited by a jumping function (that is,
the size of the approximated set of relevant nodes).
For realistic queries
(Q01-Q09  with the exception of Q08), the number of selected nodes is
more than 10\% of the number of visited nodes (this ratio is given at
Line~\textbf{(5)}).
Of particular interest is Q05. For such a query, and while the
automaton is given in an alternating and non-deterministic way, we end
up touching exactly the number of relevant nodes (the top-most
\texttt{listitem}s and the \texttt{keyword}s below them). This number
can be contrasted with the total number of nodes (more than 5
millions), most of which are completely ignored by the evaluation
function.

Line~\textbf{(3)} shows also that for a non-jumping algorithm,
our evaluation function skips, when possible, a large number of
subtrees. Of course it is necessary to traverse the whole document as
soon as a top-level ``//'' is present. 

The automata logic is better highlighted by looking at
Line~\textbf{(2)} for query Q10 to Q15. Here, it is clear that
predicates are efficiently checked. For Q11, Q12 and
Q13, the predicate check is done together with the accumulation of
\texttt{keyword} nodes, and no extra relevant node is touched. For
query Q14 and Q15, only a small number of nodes (1 and 2 respectively)
are touched in order to satisfy the predicate. Of course, the
predicate need
not be applied to root node, such optimizations are performed for any
kind of conditions, regardless of their position in the query (it is
easier to illustrate them on the single root element).

Lastly, Line~\textbf{(4)} represents the number of entries added to
the memoization table, or equivalently the number of nodes for which
the evaluation function paid a $|Q|$ factor (whereas all the others
consisted of a constant-time look-up). For practical queries, the size of
such tables is very small and the speed-up they generate is worth the
small memory overhead (a few kilo-bytes at most).

\begin{figure}
\vspace{-4mm}
\begin{minipage}{3.5cm}
\includegraphics[width=3.5cm]{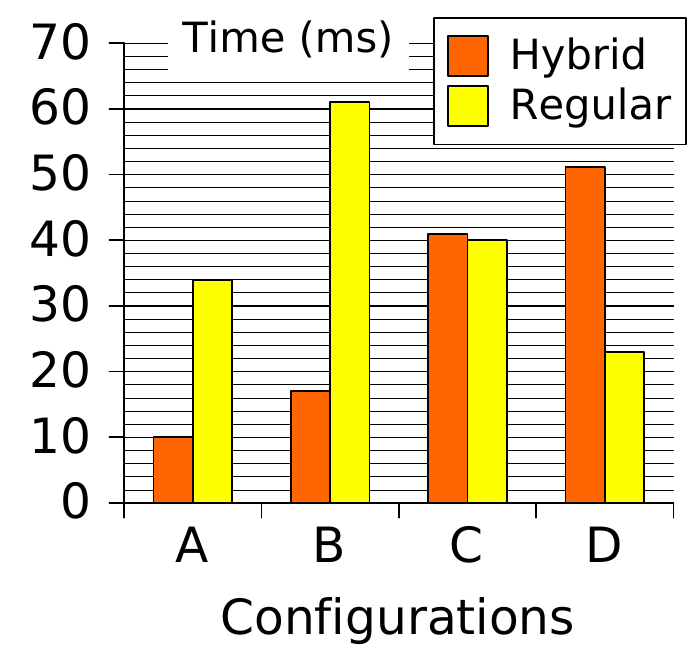}
\end{minipage}
\begin{minipage}{4cm}
\begin{tabular}{|@{~}l@{~}|@{~}c@{~}|@{~}c@{~}|@{~}c@{~}|@{~}c@{~}|}
\hline
 & \textbf{A} & \textbf{B} & \textbf{C} & \textbf{D}\\
\hline
\textbf{(1)} & 4 & 4 & 65831 &  15074 \\
\textbf{(2)} & 9 & 11 & 74302 &  33041 \\
\textbf{(3)}& 70028 & 134247 & 74302 & 35045 \\
\hline
\end{tabular}\\
\textbf{(1)} number of selected nodes\\
number of nodes visited by:\\
\hspace*{4pt}\textbf{(2)} an hybrid run\\
\hspace*{4pt}\textbf{(3)} a top-down+bottom-up run\\
\end{minipage}\\
\begin{small}
\begin{myitemize}
\item[\textbf{A}]: 75021 \texttt{listitem}, 3 \texttt{keyword} below
  \texttt{listitem}s (3 in total) and 4 \texttt{emph}s below those 3
  \texttt{keyword}s;
\item[\textbf{B}]: 75021 \texttt{listitem}, 60234 \texttt{keyword} below
  \texttt{listitem}s (60234 in total) and 4 \texttt{emph}s below those
  \texttt{keyword}s;
\item[\textbf{C}]: 9083 \texttt{listitem}, one \texttt{keyword} 
  below \texttt{listitem}s (40493 in total) and 65831 \texttt{emph}s below one
  of the \texttt{keyword} below a listitem;
\item[\textbf{D}]: 20304 \texttt{listitem}, 10209 \texttt{keyword} below
  one \texttt{listitem} (10209 in total) and 15074 \texttt{emph}s below one
  of those \texttt{keyword}.
\end{myitemize}
\end{small}
\vspace{-4mm}
\caption{Selected and visited nodes for the hybrid and top-down evaluation
  procedures, for query \texttt{//listitem//keyword//emph}}
\label{fig:hyb}
\vspace{-5mm}
\end{figure}
\noindent\textbf{Hybrid Traversal} Figure~\ref{fig:hyb} describes
the behaviour of the hybrid evaluation function for four particular
configurations of XMark documents that we manually created.

\noindent We consider the query \mbox{//{\tt listitem}//{\tt keyword}//{\tt   emph}}
and change the proportion and placement of the \texttt{listitem},
\texttt{keyword} and \texttt{emph} elements. For each such
configurations (\textbf{A} to \textbf{D}), we report the query
evaluation time for an hybrid run and for a regular
top-down+bottom-up run. We also report the number of nodes selected by
the query and the number of nodes visited by both strategies.
 Configuration \textbf{A} and \textbf{B} represent the best cases for
 the hybrid traversal:  one of the label in the query has a very
 low global count. In \textbf{A}, the count of \texttt{keyword} nodes
 is small, the evaluation starts at these nodes, checks in a pure
 bottom-up fashion that they have a \texttt{listitem} ancestor and
 collect their \texttt{emph} descendants. For configuration
 \textbf{B}, the hybrid run actually performs a pure bottom-up run of the
 query, starting at \texttt{emph} nodes. Both visit very few nodes
 compared to the relevant nodes approximated by the top-down+bottom-up
 evaluation
 (Line~\textbf{(3)}). Configuration \textbf{C} represents a case where
 the hybrid behaves like the top-down+bottom-up run, since the global
 count of \texttt{keyword} elements is low. Lastly, Configuration
 \textbf{D} is the worst-case scenario, where \texttt{keyword} as the
 lowest global count, but which is close to the number of
 \texttt{listitem} elements. Even though the top-down+bottom-up visits
 more nodes, it is twice as fast thanks to its use of jumping
 primitives. While this particular experiment seems artificial,
 configuration \textbf{A} and \textbf{B} actually simulate the
 behaviour of text-oriented queries, where the text predicate is often
 very selective. Such queries where investigated in
 \cite{DBLP:journals/corr/abs-0907-2089}, where the same hybrid
 procedures yields significant improvement over state of the art text-aware
 XPath engines.


\section{Conclusion}
\label{sec:conclu}
We have presented an effective way to reduce the number of nodes traversed
during the evaluation of a navigational XPath query, using the
novel notion of \emph{relevant nodes} for an automaton. We have shown that
this notion, coupled with a wide range of implementation techniques
made alternating selecting tree automata a compilation target of
choice for XPath queries, yielding execution speed on par with the
best XPath engines available.
While we have only focused our presentation on forward Core XPath, our
prototype actually implements backward axes (by adding ``up-moves'' to
formulas of the ASTA which are rewritten into down moves on-the-fly)
and XPath 1.0 functions. Unfortunately  ``up-moves'' are
not part of the theory and present two problems. The first one is that
we do not have yet a sound  approximation of relevant nodes in the
presence of up-move (therefore we cannot jump). The second, more
troublesome one is that with the presence of up-moves, a single
top-down followed by a bottom-up pass is not sufficient in general,
one needs an extra top-down pass (as observed by Koch in
\cite{DBLP:conf/vldb/Koch03}), or require more book-keeping operations
in the result sets. XPath 1.0 functions are naively treated as
black-boxes which are called during formula evaluation. This
defeats some of the automata optimizations since a query ``//{\tt a}[ count(.//{\tt b}) ]//c''
gets compiled into three separate automata.

As future work, we plan to generalizes the top-down approximation to
backward axes (it seems possible since ASTAs are known to not gain any
expressive power with the addition of up-moves),
extend the work in \cite{DBLP:journals/corr/abs-0907-2089} to not only
handle efficiently text predicates but also numeral predicates,
context dependent functions (e.g. ``\texttt{position()}'') and
data joins.

\bibliographystyle{plain}
\selectfont
\bibliography{main}
\appendix
\section{Selecting Tree Automata}
\label{ap:auto}
We consider an example of a 
BDSTA for which there is no equivalent
top-down deterministic STA.
\begin{example}
\label{ex:ssafssb}
Let $\mathcal{A}_{\text{//{\tt a}[.//{\tt b}]}}=\defaultAuto$ where
$\Sigma=\{\mathtt{a},\mathtt{b},\mathtt{c}\}$,
$Q=\{q_0, q_1\}$, 
$\mathcal{T}=\{q_0,q_1\}$,
$\mathcal{B}=\{q_0\}$,
$\mathcal{S}=\{(q_1,\mathtt{a})\}$,
and $\delta$ consists of the following eight transitions.
A transition $(q,x,q',q'')\in\delta$ is now written in 
the form $q\leftarrow x,(q',q'')$.
Let $\_$ denote any state in $\{q_0,q_1\}$.\\
\centerline{\small$
\begin{array}{l@{}c@{}r}
q_1 & \leftarrow & \{\mathtt{b}\}, (q_0,\_) \\
q_0 & \leftarrow & \Sigma\setminus\{\mathtt{b}\}, (q_0,\_)\\
\end{array}~~~~
\begin{array}{l@{}c@{}r}
q_1 & \Leftarrow & \{\mathtt{b}\}, (q_1,\_) \\
q_1 & \leftarrow & \Sigma\setminus\{\mathtt{b}\}, (q_1,\_) \\
\end{array}
$}
\end{example}
The automaton $\mathcal{A}_{\text{//{\tt a}[.//{\tt b}]}}$ accepts the
set of all trees:
$\LANG{\mathcal{A}_{//{\tt a}[.//{\tt b}]}}=T(\Sigma)$.
Moreover, $\mathcal{A}_{\text{//{\tt a}[.//{\tt b}]}}$ 
is a bottom-up complete BDSTA. It 
selects all the \texttt{a}-nodes that
have a \texttt{b}-node in their left subtree.
In terms of XML, this automaton realizes the XPath 
query //{\tt a}[.//{\tt b}].
We claim that there is no top-down deterministic STA 
equivalent to $\mathcal{A}_{\text{//{\tt a}[.//{\tt b}]}}$ of
Example~\ref{ex:ssassb}.
Intuitively, the top-down automaton does not know whether
or not to select an \texttt{a}-node, because this depends on the
left subtree of that node, which has not yet been processed
by the automaton.
\begin{definition}[Reachable state]
If a state $q'$ appears in the right-hand side of a rule
with $q$ in its left-hand side, then we say that
$q$ \emph{one-step reaches} $q'$, denoted by $\reach{\mathcal{A}}{q}{q'}$.
We denote by $\reachstar{\mathcal{A}}{}{}$ the
reflexive transitive closure of $\reach{\mathcal{A}}{}{}$,
and say that $q$ \emph{reaches} $q'$ if 
$q\reachstar{\mathcal{A}}{}{} q'$.
\end{definition}

We give the formal definition for the notation $\mathcal{A}[q]$ of the
restriction of an automaton to a state.
\begin{definition}[Restriction of an automaton to a set of state]
Let $\mathcal{A}=\defaultAuto$ and  $\{q_1,\dots,q_n\}\subseteq Q$,
the \emph{restriction of $\mathcal{A}$ to 
$\{q_1,\dots,q_k\}$}
is the STA
$$\mathcal{A}[q_1,\ldots,q_n]=
(\Sigma,Q',\mathcal{T}',\mathcal{B}',\mathcal{S}',\delta')$$
where $\mathcal{T}'=\{q_1,\dots,q_n\}$, 
$Q'$ is the set of the states reachable from $\mathcal{T}'$, i.e.,
$Q'=\{q' \in Q\mid\exists
q\in\mathcal{T}',~\reachstar{\mathcal{A}}{q}{q'}\}$,
and $\mathcal{B}'$, $\mathcal{S}'$, and $\delta'$ are the
restrictions to the states in $Q'$ 
of $\mathcal{B}$, $\mathcal{S}$, and $\delta$, i.e.,
$\mathcal{B}'= \mathcal{B}\cap Q'$,
$\mathcal{S}'= \{ (q,l)\in\mathcal{S}\mid q\in Q' \}$,
and
$\mathcal{\delta}'=\{(q,L,q_1,q_2)\in\delta\mid q\in Q'\}$.
\end{definition}

\subsection{Relating STAs to Ordinary Tree Automata}

In the next section we will characterize, for a given tree
$t\in T(\Sigma)$, the nodes of $t$ that are ``relevant'' for
a the STA $\mathcal{A}$. Intuitively, a node is relevant if 
$\mathcal{A}$ changes its state at that node.
However, it can be that the STA $\mathcal{A}$ is 
``badly programmed'' and changes its states at more places
than is actually necessary for the query.
Were therefore want to consider the \emph{minimal} automaton
$\mathcal{A}'$ that is equivalent to $\mathcal{A}$, where
minimal means with the least number of states.
It is well-known that for every ordinary deterministic tree automaton (TA)
there is an equivalent unique minimal one, and that it can be
computed in quadratic time.
Instead of inventing and proving a new minimization procedure
for STAs we prefer to encode them into ordinary tree automata
in such a way that the encoding allows us to obtain a minimal
STA from the minimal encoded automaton. Thus, we reduce minimization
for STAs to minimization for ordinary tree automata.

We require that the STA $\mathcal{A}$ is either
top-down or bottom-up complete.
To encode an STA into a TA, we simply 
encode the selection of a node through special labels.
We define the alphabet 
$\widehat{\Sigma}=\{\hat{\sigma}\mid \sigma\in \Sigma\}$.
Now, if $\mathcal{A}$ selects a node in a given tree (with label $l$), 
then the TA $\widehat{\mathcal{A}}$
\emph{associated to $\mathcal{A}$} accepts a tree that
has the label $\hat{l}$ at that node.
Formally, 
\[
\widehat{\mathcal{A}}=(\Sigma\cup\widehat{\Sigma},Q,
\mathcal{T},\mathcal{B},\varnothing,\widehat{\delta})
\]
where $\hat{\delta}$ is defined as follows.
Every transition $(q,L,q_1,q_2)\in\delta$ such that
there exists an $l\in L$ with $(q,l)\in\mathcal{S}$ is
changed into the new transition
$(q,L',q_1,q_2)$ of $\widehat{\mathcal{A}}$ where
$L'=\{l\in L\mid (q,l)\not\in\mathcal{S}\}$
(if $L'=\varnothing$ then the transition is removed),
and additionally we add the new transition
$(q,\widehat{L},q_1,q_2)$ to $\hat{\delta}$ where
$\widehat{L}=\{\hat{L}\mid l\in L, (q,l)\in\mathcal{S}\}$.
Finally, we make the automaton obtained so far
complete: for every $q\in Q$ let 
$L(q)=\{\sigma\in\Sigma\cup\widehat{\Sigma}\mid
\hat{\delta}(q)\not=\emptyset\}$ and, if
$L(q)\not=\varnothing$ then add the transition
$(q,L(q),q_\bot,q_\bot)$ to $\hat{\delta}$.
For the new sink state $q_\bot$ we add the transition
$(\hat{q}_\bot,\Sigma\cup\widehat{\Sigma},\hat{q}_\bot,\hat{q}_\bot)$
to $\hat{\delta}$.
It should be clear that 
\begin{enumerate}
\item[(1)] for every $t\in\mathcal{L}(\mathcal{A})$ there
exists a tree $t'\in\widehat{\mathcal{A}}$ which is obtained
from $t$ by changing the label of every $\pi\in\mathcal{A}(t)$
into $\hat{l}$, where $l=t(\pi)$.
\item[(2)] for every $t'\in\mathcal{L}(\widehat{\mathcal{A}})$ there
exists a tree $t\in\widehat{\mathcal{A}}$ obtained
by removing all hats, and, every node $\pi$ in $t'$ that
has a hat, $\pi$ is in $\mathcal{A}(t)$
\end{enumerate}
If (1) and (2) hold for two automata
$\mathcal{A}$ and $\hat{\mathcal{A}}$ then we say
that they are \emph{equivalent}, denoted by
$\mathcal{A}\equiv\hat{\mathcal{A}}$.

\begin{example}
\label{ex:recog}
The recognizer associated with the STA defined in
Example~\ref{ex:ssassb} is:
$$
\widehat{\mathcal{A}} = (\Sigma\cup\widehat{\Sigma},\{\hat{q_0},
\hat{q_1},\hat{q_\bot}\},\{
\hat{q_0} \}, \{\hat{q_0}, \hat{q_1}\}, \varnothing, \hat{\delta})
$$
where $\hat{\delta}$ is defined as:
$$
\begin{array}{l@{}c@{}l}
\hat{q_0},\{\mathtt{a}\} & \rightarrow & (\hat{q_1},\hat{q_0}) \\
\hat{q_0},\Sigma\setminus\{\mathtt{a}\} & \rightarrow & (\hat{q_0},\hat{q_0}) \\
\hat{q_0},\widehat{\Sigma} & \rightarrow &
(\hat{q_\bot},\hat{q_\bot})\\
\end{array}~
\begin{array}{l@{}c@{}l}
\hat{q_1},\{\hat{\mathtt{b}}\}\cup\Sigma\setminus\{\mathtt{b}\} & \rightarrow & (\hat{q_1},\hat{q_1}) \\
\hat{q_1},\{\mathtt{b}\}\cup\widehat{\Sigma}\setminus\{\hat{\mathtt{b}}\} & \rightarrow & (\hat{q_\bot},\hat{q_\bot}) \\
\hat{q_\bot},\Sigma\cup\widehat{\Sigma} & \rightarrow & (\hat{q_\bot},\hat{q_\bot})\\
\end{array}$$
\end{example}

The connection between an STA and its associated recognizer is
quite strong, as we state in the following lemma.

\begin{lemma}
\label{lem:markrecogequiv}
Let $\mathcal{A}$ and $\mathcal{A}'$ be two STAs, defined over the
same alphabet $\Sigma$. Then
$\mathcal{A}\equiv\mathcal{A}'$ if and only
if $\LANG{\widehat{\mathcal{A}}} = \LANG{\widehat{\mathcal{A}'}}$.
\end{lemma}

We have seen how to translate an STA into an ordinary tree
automaton. It should be clear that this translation preserves
determinism.
The translation is invertible: for any $\widehat{\mathcal{A}}$
automaton, one can build an equivalent (in the sense of
Lemma~\ref{lem:markrecogequiv}) ordinary tree automaton 
$\mathcal{A}$. However, this
inverse translation \emph{does not} preserve determinism. Indeed,
while both formalisms are equally expressive, they do not have the
same behaviour. The automaton $\widehat{\mathcal{A}}$ only needs to verify
that a tree in $T(\Sigma\cup\widehat{\Sigma})$ is in its
language. This can always be done in a bottom-up deterministic way
(it is folklore that bottom-up tree automata can be determinized,
see~\cite{tatabook}).

For our purpose, it is enough to observe that 
if a deterministic automaton $\widehat{\mathcal{A}}$ is
``selecting-unambiguous'', then it can be transformed
into a deterministic SA.
Formally, the tree automaton 
$\mathcal{A}=
(\Sigma\cup\widehat{\Sigma},Q,\mathcal{T},\mathcal{B},\varnothing,\delta)$ 
is \emph{selecting-unambiguous} if and only if
for every $q\in Q$, and for every $t\in\LANG{\mathcal{A}[q]}$:
\begin{itemize}
\item if $t(\epsilon) = \sigma\in\Sigma$, then $t[ \epsilon \leftarrow \hat{\sigma}]\notin\LANG{\mathcal{A}[q]}$
\item if $t(\epsilon) = \hat{\sigma}\in\widehat{\Sigma}$, then $t[ \epsilon \leftarrow \sigma]\notin\LANG{\mathcal{A}[q]}$
\end{itemize}

\begin{lemma}
Let $\mathcal{A}$ be a complete TA. Then $\widehat{\mathcal{A}}$ is
selecting-unambiguous.
\end{lemma}
\begin{lemma}
\label{def:recogtomark}
Let $\mathcal{A}'$
be a complete selecting-unambiguous TDTA (resp. BDTA). 
There effectively exists a complete TDSTA (resp. BDSTA)
$\mathcal{A}$ such that $\hat{\mathcal{A}}\equiv\mathcal{A}'$. 
\end{lemma}
\begin{proof}(sketch)
The proof builds the automaton $\mathcal{A}$ as such. For each transition
$(q,L,q_1,q_2)\in\delta'$. We split the transition in two,
$(q,L',q_1,q_2)\in\delta'$ and
$(q,L'',q_1,q_2)\in\delta'$ where $L'= L \cap \Sigma$ and $L'' =
L\cap\widehat{\Sigma}$ (if $L'$ or $L''$ is empty, we just skip it).
Since $\mathcal{A}'$ is marking-unambiguous, if $\sigma\in L'$, then
$\hat{\sigma}\notin L''$ (and vice versa). If neither $q_1$ nor $q_2$
is a sink state, then we add
$(q,L',q_1,q_2)\in\delta'$ as a transition to $\delta$ and if 
$L''=\{\hat{\sigma_1},\ldots,\hat{\sigma_k}\}$ we add
$(q,\{\sigma_1,\ldots,\sigma_k\},q_1,q_2)$ to $\delta$ and
$(q,\sigma_i)$ to $\mathcal{S}$.
Once this is done for all transitions, we remove all unreachable
states and we obtain $\mathcal{A}$.
\end{proof}

Now that we have established a precise correspondence between STAs
and TAs we get for free some properties of TAs,
such as minimization.

\subsection{Minimization}
\label{ap:mini}
As mentioned before, minimimal here means, the smallest number
of states.
Given a BDTA 
$\mathcal{A}=(\Sigma,Q,\mathcal{T},\mathcal{B},\delta)$, the standard
algorithm for minimization (see, e.g.,~\cite{tatabook})
builds the set of equivalence classes for every state in $Q$. 
Two states $q$ and $q'$
are in the same equivalence class if and only if
$\LANG{\mathcal{A}[q]} = \LANG{\mathcal{A}[q']}$. The algorithm
initializes the set of equivalence classes with
$E_0=\{Q\setminus\mathcal{T}, \mathcal{T}\}$.
The intuition is that final and non-final states are not in the same
equivalence classes (indeed, if
$q$ is a final state and $q'$ not a final state, then $\mathcal{A}[q]$
accepts the null tree $\#$ while $\mathcal{A}[q']$ does not, hence
$\LANG{\mathcal{A}[q]} \neq \LANG{\mathcal{A}[q']}$).
The algorithm proceeds then to refine the equivalence relation.
We note $q~E_n~q'$ the fact that $q$ and $q'$ are equivalent in the
equivalence relation $E_n$, that is there exists $S\in E_n$ such that
$q\in S$ and $q'\in S$. From $E_n$ the algorithm computes a finer
equivalence relation $E_{n+1}$ such that $q~E_{n+1}~ q'$ if:
\begin{itemize}
\item $q~E_n~q'$; 
\item $\forall \sigma\in\Sigma,\forall q_1,q_2\in Q \delta(q1,q,l) =
  \delta(q1,q',l)$ and $\delta(q,q_2,l) = \delta(q',q2,l)$.
\end{itemize}
The procedures stops when $E_n = E_{n+1}$. The case of TDTA is
similar.

Of course we would like, given a selecting automaton $\mathcal{A}$, to
compute is associated recognizer $\widehat{\mathcal{A}}$, minimize it using
the standard procedure and translate it back into a selecting automaton.
However, as we have seen, translating a recognizer into a selecting
automaton does not always preserve determinism. Fortunately, we can
show that the property of selecting unambiguousness is preserved by the
minimization procedure.
\begin{lemma}
\label{lem:minimnonambi}

Let $\widehat{\mathcal{A}}$
be a complete TDTA (resp. BDTA) over the alphabet
$\Sigma\cup\widehat{\Sigma}$.
Let $\widehat{\mathcal{A}}_{\textit{min}}$ be the minimal automaton
such that $\LANG{\widehat{\mathcal{A}}_{\textit{min}}} =
\LANG{\widehat{\mathcal{A}}}$. 
If $\widehat{\mathcal{A}}$ is selecting-unambiguous, then
so is $\widehat{\mathcal{A}}_{\textit{min}}$.
\end{lemma}
\begin{proof}
First let us remark than since $\widehat{\mathcal{A}}$ is
selecting-unambiguous, then $\forall q\in \widehat{Q}, \forall t\in 
\LANG{\widehat{\mathcal{A}}[q]}$, if $t(\epsilon)=\sigma\in\Sigma$
then $t[\epsilon \leftarrow \hat{\sigma}]\notin
\LANG{\widehat{\mathcal{A}}[q]}$ and if
$t(\epsilon)=\hat{\sigma}\in\widehat{\Sigma}$
then $t[\epsilon \leftarrow \sigma]\notin
\LANG{\widehat{\mathcal{A}}[q]}$.

Now suppose that there are two states $q_1,q_2\in\widehat{Q}$ such that 
$\exists \sigma(t_1,t_2)\in \LANG{\widehat{\mathcal{A}}[q_1]}$ and
$\exists \hat{\sigma}(t_1,t_2)\in
\LANG{\widehat{\mathcal{A}}[q_2]}$. $\widehat{\mathcal{A}}_{\textit{min}}$
is selecting-unambiguous if and only if $q_1$ and $q_2$ are not in the
same equivalence class (if they where, then there would be a state
in $q'\in Q_{\textit{min}}$ for which the selecting-unambiguous
property do not hold, the state representing the equivalence class
of $q_1$ and $q_2$).
We must therefore show that
$\LANG{\widehat{\mathcal{A}}[q_1]}\neq\LANG{\widehat{\mathcal{A}}[q_2]}$
This is immediate: since $\mathcal{A}$ is selecting unambiguous,
and since $\sigma(t_1,t_2)\in \LANG{\widehat{\mathcal{A}}[q_1]}$,
then $\hat{\sigma}(t_1,t_2)\notin\LANG{\widehat{\mathcal{A}}[q_1]}$.
However
$\hat{\sigma}(t_1,t_2)\in\LANG{\widehat{\mathcal{A}}[q_2]}$ and
therefore
$\LANG{\widehat{\mathcal{A}}[q_1]}\neq\LANG{\widehat{\mathcal{A}}[q_2]}$. 
\end{proof}

Using this lemma, we can state the existence of a minimal selecting tree
automaton.
\begin{theorem}
\label{thm:minim}
Let $\mathcal{A}$ be a complete TDSTA (resp. BDSTA). There
effectively exists a complete TDSTA (resp. BDSTA) 
$\mathcal{A}_{\textit{min}}$ which is equivalent to
$\mathcal{A}$ and no other equivalent TDSTA (resp. BDSTA) has
less states than $\mathcal{A}_{\textit{min}}$.
\end{theorem}

Theorem~\ref{thm:minim} states the existence of a minimal selecting
automaton and also give a way to compute it. Indeed, it is sufficient
to  translate a selecting automaton into a recognizer,
minimize the
latter and  transform it back into a selecting automaton. However, the
proof of Lemma~\ref{lem:minimnonambi} hints us toward a more direct
method. Indeed in a recognizer, if a state $\hat{q}_1$ accepts some tree
$\sigma(t_1,t_2)$ and a state $\hat{q}_2$ accepts the tree
$\hat{\sigma}(t_1,t_2)$, then $\hat{q}_1$ and $\hat{q}_2$ are in different
equivalence classes. In the transformation from recognizer to selecting
automaton, $q_2,\sigma$ becomes a selecting configuration. Therefore, if
two states $q_1$ and $q_2$ are such that $q_1,\sigma\notin\mathcal{S}$
and $q_2,\sigma\in\mathcal{S}$ then these two states are not in the
same equivalence class. Minimizing an selecting automaton can therefore
be achieved by using the standard algorithm, but where the initial
relation $E_0$ is:
$$
E_0 = \{ \begin{array}{l}
  \{ q\in Q\mid q\in\mathcal{F},q\in\mathcal{S} \},\\
  \{ q\in Q\mid q\in\mathcal{F},q\notin\mathcal{S} \},\\
  \{ q\in Q\mid q\notin\mathcal{F},q\in\mathcal{S} \},\\
  \{ q\in Q\mid q\notin\mathcal{F},q\in\mathcal{S} \} 
\end{array} \}.
$$
Here $\mathcal{F}$ stands for the set of final states, that is
$\mathcal{T}$ for BDTAs and $\mathcal{B}$ for TDTAs.
\section{Relevance}
\subsection{Top-Down Relevance}

\begin{algorithm}[Top-down traversal with jumping]~\\[-4mm]
\label{alg:tdj}
\begin{small}
  \begin{description}
  \item[Input:] Minimal TDTA
    $\mathcal{A}=(\Sigma,Q,\mathcal{F},\mathcal{I},\mathcal{S},\delta)$
    and a tree $t$
  \item[Output:] (possibly empty) Mapping from nodes of $t$ to
    states of $\mathcal{A}$.

  \end{description}
  \begin{lstlisting}
    let following($\pi,L,\pi_0$)= $\label{code:following}$ 
       if $\pi$ = $\Omega$ then return $\varnothing$ 
       else return $\{ \pi \}\cup$ following($\mathbf{f}_t(\pi,L,\pi_0)$,$L$,$\pi_0$);

    let relevant_nodes ($t,\pi,q$) = $\label{code:relnodes}$
        if $\exists L\subset \Sigma, (q,L,q,q) \in \delta$ and $\neg$is_marking($q$)
        then { $L'$ := $\Sigma \setminus L$; $\label{code:tdj_casei}$
           if $t(\pi) \in L'$ then return $\{ \pi \}$;
           $\pi'$ := $\mathbf{d}_t(\pi,L')$;
           return $\{ \pi' \} \cup$ follow($\pi'$,$L'$, $\pi$)
        } else
        if $\exists L\subset \Sigma, (q,L,q,q_\top) \in \delta$
           and is_universal($q_\top$) and $\neg$is_marking($q$)
        then { $L'$ := $\Sigma \setminus L$;
           if $t(\pi) \in L'$ then return $\{ \pi \}$;
           $\pi'$ := $\mathbf{l}_t(\pi,L')$;
           if $\pi'$ = $\Omega$ then return $\varnothing$ else return $\{ \pi' \}$ 
        } else
        if $\exists L\subset \Sigma, (q,L,q_\top,q) \in \delta$
           and is_universal($q_\top$) and $\neg$is_marking($q$)
        then { $L'$ := $\Sigma \setminus L$;
           if $t(\pi) \in L'$ then return $\{ \pi \}$;
           $\pi'$ := $\mathbf{l}_t(\pi,L')$;
           if $\pi'$ = $\Omega$ then return $\varnothing$ else return $\{ \pi' \}$
        } else
        return $\{ \pi \}$;

    let td_jump_rec ($\pi,q$) = $\label{code:top_down_jump_rec}$
     $l$: = $t(\pi)$;
     if $l = \#$ then $\label{code:tdj_basic}$
       if $q\in \mathcal{B}$ then return $\{ \pi \mapsto q \}$
       else throw $\texttt{Failure}$
     else {
     $\{ q_1, q_2 \}$ := $\delta(q,l)$;
     if is_sink($q_1$) or is_sink($q_2$) then throw $\texttt{Failure}$;
     lnodes := relevant_nodes ($t,\pi\cdot 1,q_1$);
     rnodes := relevant_nodes ($t,\pi\cdot 2,q_2$);
     return $\displaystyle \{ \pi \mapsto q \} \cup \bigcup_{\pi_1\in \textit{lnodes}}$topdown_jump_rec($\pi_1,q_1$)
                      $\displaystyle\cup\bigcup_{\pi_2\in \textit{rnodes}}$topdown_jump_rec($\pi_2,q_2$);
     }
    
    let topdown_jump($t$,$(\Sigma,Q,\mathcal{F},\{ q \},\mathcal{S},\delta)$) = $\label{alg:tddef}$
     try {
      nodes := relevant_nodes ($t,\epsilon,q$);
      return $\displaystyle\bigcup_{\pi\in \textit{nodes}}$topdown_jump_rec($\pi,q$);
     } $\texttt{catch}$ ($\texttt{Failure}$) { return $\varnothing$ }
  \end{lstlisting}
\end{small}
\end{algorithm}

\subsection{Bottom-Up Relevance}
\label{ap:bu}

\begin{small}
\begin{algorithm}[Bottom-up evaluation]~\\[-4mm]
      \begin{description}
      \item[Input]: A BDTA $\mathcal{A} = \{ \Sigma, Q, \mathcal{T},
        \{ q_0 \}, \mathcal{S}, \delta\}$ a tree $t$ a sequence
        $$S_0 = 
        (\pi_0,q_0)\texttt{, }(\pi_1,q_0)\texttt{,
        }\ldots\texttt{, }(\pi_n,q_0)
        $$
        where the $\pi_i$ are the leaves of $t$ in pre-order.
      \item[Output]: A mapping from nodes of $t$ to states of $\mathcal{A}$
      \end{description}
\begin{lstlisting}
    let bottom_up_rec $(S,t,R)$ = 
      $\texttt{switch}$ $S$ {
      case $(\pi,q)$:
       if $\pi$ = $\epsilon$ and $q\in\mathcal{T}$ 
       then return $\{ \epsilon\mapsto q \}\cup R$,();
       else throw $\texttt{Failure}$;

      case $(\pi_1,q_1)$,$(\pi_2,q_2)$,$S'$:
       if siblings($\pi_1,\pi_2$)
       then {
          $\pi$ := parent $\pi_1$;
          $\{ q \}$ := $\delta(q_1,q_2,t(\pi))$;
       return bottom_up_rec($((\pi,q),S'), t,\{\pi \mapsto q\}\cup R$)
       } else { 
          $R'$,$S''$ := bottom_up_rec($((\pi_2,q_2),S'),t,R$);
          return bottom_up_rec ($((\pi_1,q_1),S''),t,R'$);
     case ():
          return $R$,();
       }

     let bottom_up ($t, S_0, \mathcal{A}$) = 
       try {
         $R$,_ := bottom_up_rec($S_0,t,\{ \pi\mapsto q\mid (\pi,q)\in S_0\}$);
         return $R$;
       } catch ($\texttt{Failure}$)
            return $\varnothing$;
      \end{lstlisting}
    \end{algorithm}
\end{small}

\begin{figure}
\begin{center}
\includegraphics[width=6cm]{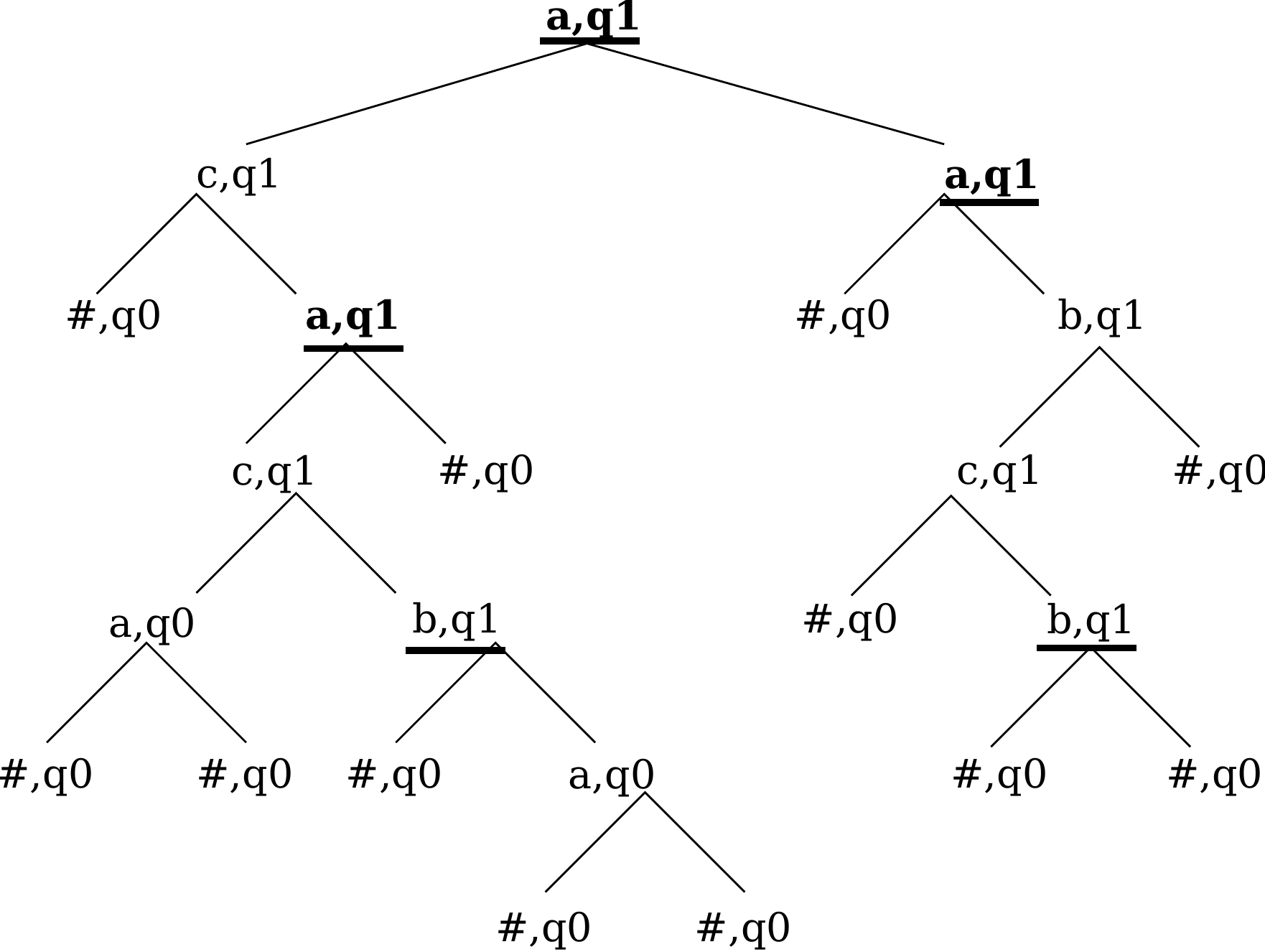}
\end{center}
\caption{Bottom-up run of automaton $\mathcal{A}$ from
  Example~\ref{ex:bu1}}
\label{fig:buex1}
\end{figure}

\begin{example}
\label{ex:bu1}
$$\mathcal{A}_{\text{//\tt{a}[.//{\tt b}]}}=(\underset{\Sigma}{\{\texttt{a},\texttt{b},\texttt{c}\}},\underset{Q}{\{q_0,q_1\}},\underset{\mathcal{T}}{\{q_0\}},\underset{\mathcal{B}}{\{q_0,q_1\}},\underset{\mathcal{S}}{\{(q_1,\texttt{a})
  \}},\delta )$$
A transition $(q,L,q',q'')\in\delta$ is written in 
the form \mbox{$q\leftarrow L(q',q'')$} and the wildcard  $\_$ denotes any
state in $\{q_0,q_1\}$. $\delta$ is defined by:
$$
\begin{array}{l@{}c@{}r}
q_1 & \leftarrow & \{\mathtt{b}\}, (q_0,\_) \\
q_0 & \leftarrow & \Sigma\setminus\{\mathtt{b}\}, (q_0,\_)\\
\end{array}~~~~
\begin{array}{l@{}c@{}r}
q_1 & \Leftarrow & \{\mathtt{a}\}, (q_1,\_) \\
q_1 & \leftarrow & \Sigma\setminus\{\mathtt{a}\}, (q_1,\_) \\
\end{array}
$$
\end{example}
A run of this automaton on an input tree is given in
Figure~\ref{fig:buex1}.
This automaton selects all the \texttt{a}-labelled node which are
above a \texttt{b}-labelled node. The selected nodes are circled
and the relevant nodes are underlined. As in the general case and the
TDSTA case, selected nodes are relevant. Otherwise, we
can remark that any subtree whose root is in state $q_0$ contain only
non relevant nodes. In the case of minimal BDSTAs, the state $q_0$ allows to
skip subtrees (as $q_\top$ for TDSTAs). Indeed in a minimal BDSTA,
$q_0$ is the only state which accepts a null-tree $\#$. But
conversely, any subtree which is recognized in $q_0$ could be replaced
by a null-tree without changing the semantics of the query. Thus,
skipped subtrees are those whose root is in state $q_0$.
For skipping nodes along a path,  the same conditions as previously
apply: either the automaton remains in the same state for a node and
both its children, or the root and one of its children are in the same
state and the other children can be skipped, that is, is in state
$q_0$.
\section{Automata for XPath}
\label{ap:implem}
\begin{definition}[XPath fragment]
An \emph{XPath expression} is a finite production of the following
grammar, with start symbol Core:\\
\begin{small}
\begin{tabular}{lcl}
Core&::=& LocationPath $|$ `/' LocationPath\\
LocationPath&::=&LocationStep (`/' LocationStep)*\\
LocationStep&::=&Axis `::' NodeTest\\
&&$|$ Axis `::' NodeTest `[' Pred `]'\\
Pred&::=& Pred `and' Pred $|$ Pred `or' Pred\\ 
&&$|$ `not' `(' Pred `)' $|$ Core $|$ `(' Pred `)'\\
Axis&::=& \texttt{descendant} $|$ \texttt{child}\\
&  & $|$ \texttt{following-sibling} $|$ \texttt{attribute}\\
NodeTest & ::= & \emph{tag} $|$ \texttt{*} $|$ \texttt{node()} $|$ \texttt{text()}\\
\end{tabular}
\end{small}
\end{definition}
 The following
example clearly shows why using normal STAs would cause an exponential
blow-up:
\begin{example}
Consider the XPath query:\\
 \centerline{\texttt{//x[ (a$_1$ or a$_2$) and \ldots~and (a$_{2n-1}$ or a$_{2n}$) ]}}
where the \texttt{a}$_i$ are pairwise distinct
labels. The ASTA for this query is:\\
\centerline{\small$
\begin{array}{l@{}c@{}l}
  q_{\texttt{x}}, \{ x \} & \Rightarrow & (\downarrow_1
  q_{\texttt{a}_1} \lor \downarrow_1 q_{\texttt{a}_2})\land \ldots\land (\downarrow_1
  q_{\texttt{a}_{2n-1}} \lor \downarrow_1 q_{\texttt{a}_{2n}})\\
  q_{\texttt{x}}, \Sigma&\rightarrow &
  \downarrow_1 q_{\texttt{x}} \lor \downarrow_2 q_{\texttt{x}}\\
  q_{\texttt{a}_i}, \{ q_{\texttt{a}_i}\} & \rightarrow & \top\\
  q_{\texttt{a}_i}, \Sigma & \rightarrow & \downarrow_2 q_{\texttt{a}_i}\\
\end{array}
$}
\end{example}
This ASTA has: $2\cdot n +1$ states, $4\cdot n+2$ transitions, one of
length $2\cdot n$ and the other of fixed length (less than 3).
It is well known that converting this ASTA into an STA yield an
exponential blow-up (since one has to compute the disjunctive normal
form of the formulas; for the first transition, the DNF 
has size $2^n$).

\noindent\textbf{Evaluation of formulas and node selection}:
We define the notion of result sets an the semantics of the evaluation
of formulas, which also handles node selection.
\begin{definition}[Result set]
Let $\mathcal{A}=(\Sigma,\mathcal{Q},\mathcal{T},\delta)$ be an
ASTA and $t\in T(\Sigma)$. A \emph{result set} is a mapping from
states in $Q$ to sets of nodes in $\DOM{t}$. Given a mapping
$\Gamma$, we denote by $\Gamma(q)$ the set of states associated with
$q$ (the empty set if $q$ is not in $\DOM{\Gamma}$) and we define the
union of two mappings as:\\
\centerline{$(\Gamma_1\cup\Gamma_2) (q) = \Gamma_1(q) \cup \Gamma_2(q)$}
\end{definition}

We can now define the evaluation of a set of transitions for an
automaton.
\begin{definition}[Evaluation of a set of transitions]
Let\\ 
\centerline{$\mathcal{A}=(\Sigma,\mathcal{Q},\mathcal{T},\delta)$}
 be an ASTA,
$t\in T(\Sigma)$ a tree and $\textit{Trs}\subseteq\delta$ a set of
transitions. The evaluation of $\textit{Trs}$ for a node
$\pi\in\DOM{t}$ is a result set given by the function:\\
\centerline{$
\begin{array}{l}
\texttt{eval\_trans}(\Gamma_1,\Gamma_2,\pi,\textit{Trs}) =\\
\multicolumn{1}{r}{\displaystyle\bigcup_{(q,L,\rightarrow,\phi)\in\textit{Trs}}\{ q\mapsto S\mid
\Gamma_1,\Gamma_2\vdash_\mathcal{A}\phi = (\top,S)\}}\\
\multicolumn{1}{r}{\cup
\displaystyle  \bigcup_{(q,L,\Rightarrow,\phi)\in\textit{Trs}}\{ q\mapsto
  \{\pi\}\cup S\mid
\Gamma_1,\Gamma_2\vdash_\mathcal{A}\phi = (\top,S)\}}\\
\end{array}
$}
where $\Gamma_1$ and $\Gamma_2$ are result sets, and
$\Gamma_1,\Gamma_2\vdash_\mathcal{A}\phi = (b,S)$ is the judgement
derived by the rules in Figure~\ref{fig:formsem}.
\end{definition}


\newcommand{\smallfrac}[2]{\frac{\textrm{\footnotesize
      $#1$}}{\textrm{\footnotesize $#2$}}}
\begin{figure}
\begin{small}
\begin{displaymath}
\begin{array}{c}
  \smallfrac{}{
    \Gamma_1,\Gamma_2\vdash_{\mathcal{A}} \top
    = (\top,\emptyset)}\textbf{\small(true)} ~

  \smallfrac{\Gamma_1,\Gamma_2\vdash_{\mathcal{A}}\phi
    = (b,R)}
  {\Gamma_1,\Gamma_2\vdash_{\mathcal{A}} \lnot\phi
    = (\overline{b},\emptyset)}\textbf{\small(not)} \\[10pt]
  
  \smallfrac{\begin{array}{c}
      \Gamma_1,\Gamma_2\vdash_{\mathcal{A}}
      \phi_1=(b_1,\Gamma'_1)\\
      \Gamma_1,\Gamma_2\vdash_{\mathcal{A}}
      \phi_2=(b_2,\Gamma'_2)\\
      \end{array}
  }
  {\Gamma_1,\Gamma_2\vdash_{\mathcal{A}} \phi_1\lor\phi_2
    = (b_1,\Gamma'_1)\varovee (b_2,\Gamma'_2)}\textbf{\small(or)} \\[10pt]
    \smallfrac{\begin{array}{c}
      \Gamma_1,\Gamma_2\vdash_{\mathcal{A}}
      \phi_1=(b_1,\Gamma'_1)\\
      \Gamma_1,\Gamma_2\vdash_{\mathcal{A}}
      \phi_2=(b_2,\Gamma'_2)\\
      \end{array}
  }
  {\Gamma_1,\Gamma_2\vdash_{\mathcal{A}} \phi_1\land\phi_2
    = (b_1,\Gamma'_1)\varowedge (b_2,\Gamma'_2)}\textbf{\small(and)} \\[10pt]
    \smallfrac{q\in\DOM{\Gamma_i}}
  {\Gamma_1,\Gamma_2\vdash_{\mathcal{A}}
    \downarrow_i q = (\top,\Gamma(q))}\text{\small for~$i\in\{1,2\}$}~\textbf{\small(left,right)}\\[10pt]
  \smallfrac{\textrm{when no
      other rule applies}}{\Gamma_1,\Gamma_2\vdash_{\mathcal{A}}
    \phi = (\bot,\emptyset)}
\end{array}
\end{displaymath}
{\small
where:
\begin{displaymath}
\begin{array}{l}
  \overline{\top} =  \bot ~~ \overline{\bot}  =  \top\\
  (b_1,\Gamma_1)\ovee(b_2,\Gamma_2) =  \left\{\small\begin{array}{cr}
      \top,\Gamma_1 & \textrm{if $b_1 = \top$, $b_2 = \bot$}\\
      \top,\Gamma_2 & \textrm{if $b_2 = \top$, $b_1 = \bot$}\\
      \top,\Gamma_1\cup \Gamma_2 & \textrm{if $b_1 = \top$, $b_2 = \top$}\\
      \bot,\emptyset & \textrm{otherwise}\\
    \end{array}\right.\\
  (b_1,\Gamma_1)\owedge(b_2,\Gamma_2) =  \left\{\small\begin{array}{cr}
      \top,\Gamma_1\cup \Gamma_2 & \textrm{if $b_1 = \top$, $b_2 = \top$}\\
      \bot,\emptyset & \textrm{otherwise}\\
    \end{array}\right.
\end{array}
\end{displaymath}}
\end{small}
\caption{Inference rules defining the evaluation of a formula}
\label{fig:formsem}

\end{figure}

These rules are pretty straightforward and combine the rules for a
classical alternating automaton, with the rules of a marking
automaton. Rule~\textbf{(or)} and \textbf{(and)}
implements the Boolean connective of the formula and
collect the marking found in their true sub-formulas.
Rules~\textbf{(left)} and \textbf{(right)} (written as a rule
schema for
concision) evaluate to true if the state $q$ is in the corresponding
set. Intuitively, states in $\Gamma_1$ (resp. $\Gamma_2$) 
are those accepted in the left (resp. right) subtree of the input
tree. 
To
handle selection, we proceed as follows. Assuming the left subtree
returned a result set $\Gamma_1$ and the right subtree a result set
$\Gamma_2$:
\begin{myitemize}
\item[(1)] For each $q,L\Rightarrow \phi$ such that $\phi$
  evaluates to $\top$ ($\downarrow_i q'$ evaluates to $\top$ if
  $q'\in\DOM{\Gamma_i}$), add the mapping $q\mapsto \{\pi\}$ to
  $\Gamma$;
\item[(2)] For each $q,L\rightarrow\phi$ or
  $q,L\Rightarrow\phi$, for which $\phi$ evaluates to $\top$, 
  if $\downarrow_i q'\in\phi$ evaluates to $\top$, add the mapping
  $q\mapsto \Gamma_i(q')$ to $\Gamma$.
\end{myitemize}
This is done by the function \texttt{eval\_trans} 
 Informally we remember each node
which was selected by a particular transition (1) and for each
selected node in state $q'$ we propagate it to $q$ if it contributes
to the truth of a formula proving $q$. The selected nodes which gets
propagated to a state in $\mathcal{T}$ are therefore part of an
accepting run and constitute the result of the query. If we take
the example run given in Figure~\ref{fig:tdaex} of
Section~\ref{sec:implem}, node selection is performed as follows.
Consider the rightmost $c$ node in the
figure ($\star$). This node was entered in state $\{ q_0,q_1,q_2\}$,
therefore the active transitions for it are:\\
\centerline{$
\begin{array}{l}
\{   q_0, \Sigma\rightarrow \downarrow_1 q_0 \lor \downarrow_2 q_0;~~ 
q_1, \Sigma\rightarrow \downarrow_1 q_1 \lor \downarrow_2 q_1;~~
q_2, \{c\}\rightarrow \top;\\
\multicolumn{1}{r}{q_2, \Sigma\rightarrow \downarrow_2 q_2\}}\\
\end{array}
$}
and the result sets for its left and right subtrees are 
$\varnothing$ (since the calls to both left and right move failed).
In this environment only the third transition is satisfied, the result
set returned for this node is therefore $\Gamma_1=\{
q_2\mapsto\varnothing\}$. Returning from the recursive calls, we
arrive on the $b$ node above it, for which the active transitions
are:\\
\centerline{$
\begin{array}{l}
\{   q_0, \Sigma\rightarrow \downarrow_1 q_0 \lor \downarrow_2 q_0;~
q_1, \{ b\} \Rightarrow \downarrow_1 q_2;
~q_1, \Sigma\rightarrow \downarrow_1 q_1 \lor \downarrow_2 q_1;\}\\
\end{array}
$}
Evaluated under the results $(\Gamma_1,\varnothing)$ for the left and
right subtrees, only the second transition is satisfied. Furthermore,
this transition is a selecting one, it therefore returns result set
$\Gamma_2=\{ q_1\mapsto\{\pi_b\}\}$ where $\pi_b$ is the identifier of
this node. The parent of this $b$ node is again a $b$ node where the
same transitions are active. However the result sets for the left and
right subtrees are $(\varnothing,\Gamma_2)$. Under these hypothesis
only the third transition can be satisfied (and it is a not a
selecting one). The current $b$ node is therefore not selected,
but the result set is $\Gamma_3=\{ q_1\mapsto \Gamma_2(q_1)\}$
(since $\downarrow_2 q_1$ evaluated to $\top$ during the evaluation of
the third transition). We have $\Gamma_3=\{ q_1\mapsto \{\pi_b\}\}$.
We now move onto the $a$ parent of this $b$ node, where the active
transitions are:\\
\centerline{$
\{   q_0, \{a\}\rightarrow \downarrow_1 q_1;~  
   q_0, \Sigma\rightarrow \downarrow_1 q_0 \lor \downarrow_2 q_0;
\}
$}
evaluated under the assumptions $(\Gamma_3,\varnothing)$. Here the
first formula evaluates to $\top$, yielding the result set $\Gamma_4=\{
q_0\mapsto \Gamma_3(q_1)\} = \{q_0\mapsto \{ \pi_b\}\}$. We now see
that the node $\{\pi_b\}$ has been ``promoted'' to state $q_0$. Using
this technique we can ensure that nodes selected non-deterministically
during the bottom-up run are kept only if they propagate up to the
starting state $q_0$, in which case they are part of the result.
\section{Experiments}
\label{ap:experiments}
\begin{figure}
\includegraphics[width=7.3cm]{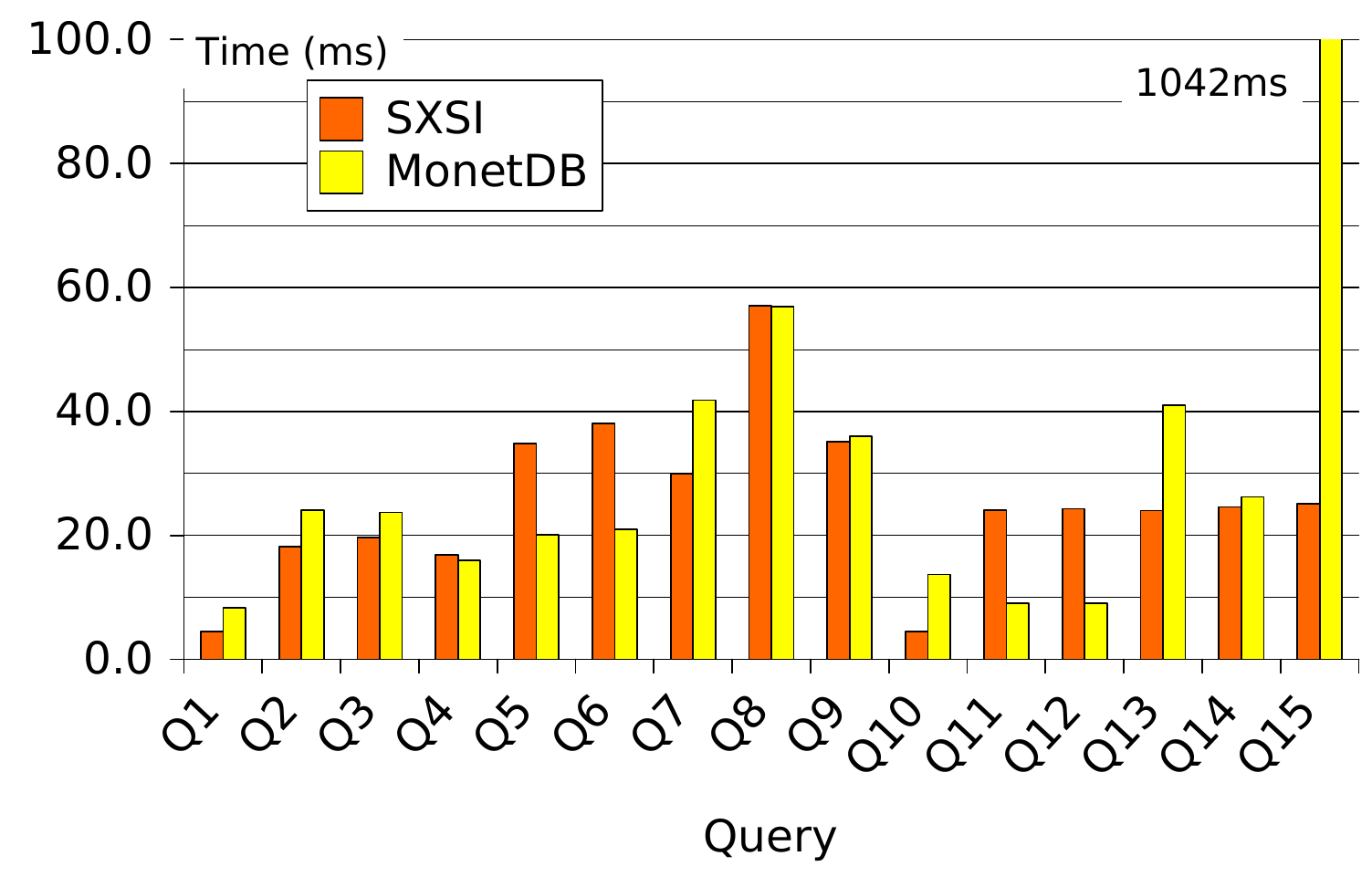}
\caption{Query answering time for the SXSI and MonetDB}
\label{fig:vsmonet}
\end{figure}

\noindent\textbf{Experimental Setup} tests were executed on an Intel
Xeon Core 2 Duo, 3 Ghz, with 4GB of RAM. We used Ubuntu Linux 9.10
distribution, with kernel 2.6.32 and 64 bits userland. Our
implementation was compiled using g++ 4.4.1 and OCaml 3.11.1. We used
version v4.34.0 of the MonetDB Server, with 32 bits OIDs.
Experimental results for query Q01~to~Q15 are given in
Figure~\ref{fig:vsmonet}. For both engines, the results was
materialized in memory but not serialized. We took the best of 5
consecutive runs for each query.


\end{document}